%% file: main.tex
\pdfoutput=1
\PassOptionsToPackage{table}{xcolor}
\documentclass[11pt]{article}

\usepackage[final]{EMNLP2023}
\usepackage{times}
\usepackage{latexsym}
\usepackage[T1]{fontenc}
\usepackage[utf8]{inputenc}
\usepackage{microtype}
\usepackage{inconsolata}
\usepackage{xcolor}

\usepackage{xspace}
\usepackage{graphicx}
\usepackage{multirow}
\usepackage{amsfonts}
\usepackage{todonotes}

\input{macro.tex}
\newcommand{\ours}[0]{\textsc{DuET}\xspace}
\title{\ours: Dual Execution for Test Output Prediction with \\ Generated Code and Pseudocode}

\author{
Hojae Han$^{1}$ Jaejin Kim$^{2,3}$ Seung-won Hwang$^{2,3}$\thanks{~~~Corresponding author.} $ $ Yu Jin Kim$^{4}$ Moontae Lee$^{4}$\\
$^{1}$ETRI, $^{2}$Seoul National University, $^3$SNU-LG AI Research Center, $^4$LG AI Research \\
\texttt{hojae.han@etri.re.kr} \texttt{\{jaejin.kim,seungwonh\}@snu.ac.kr}\\
\texttt{\{yujin.kim,moontae.lee\}@lgresearch.ai} \\
}

\begin{document}
\maketitle

\input{content/0_abstract}
\input{content/1_intro}
\input{content/2_preliminary_hj}
\input{content/3_proposed_hj}

\input{content/4_experiments}
\input{content/6_conclusion}

\bibliography{custom}
\bibliographystyle{acl_natbib}

\appendix
\input{content/Appendix}

\end{document}

%% file: macro.tex
\usepackage{siunitx}
\usepackage{listings}

\usepackage{ulem}
\usepackage{comment}
\usepackage{footmisc}
\usepackage{amsmath}
\usepackage{pifont}
\usepackage{tabularx}
\usepackage{caption, booktabs}
\usepackage{algorithm}
\usepackage{algpseudocode}
\usepackage{amssymb}
\usepackage{setspace}
\usepackage{bbm}
\usepackage{arydshln}

\newcommand{\edit}[1]{\textcolor{black}{#1}}
\newcommand{\edithj}[1]{\textcolor{black}{#1}}

\makeatletter
\newcommand\footnoteref[1]{\protected@xdef\@thefnmark{\ref{#1}}\@footnotemark}
\makeatother
\newcolumntype{P}[1]{>{\centering\arraybackslash}p{#1}}

\usepackage{xspace}
\newcommand{\ourslink}[0]{\url{https://github.com/ldilab/DuET}}
\usepackage{graphicx}
\usepackage{adjustbox}
\usepackage{subfig}
\makeatletter
\newcommand{\thickhline}{
\noalign {\ifnum 0=`}\fi \hrule height 1pt
    \futurelet \reserved@a \@xhline
}
\newcolumntype{"}{@{\hskip\tabcolsep\vrule width 1pt\hskip\tabcolsep}}
\makeatother
\usepackage{mathabx}
\usepackage{amsfonts}

\makeatletter
\newcommand*{\blackleq}{
    \mathrel{
        \mathpalette\@blackleq{}
    }
}
\newcommand*{\@blackleq}[2]{
    \vcenter{
        \m@th
        \setbox0=\hbox{$#1\mkern3mu$}
        \setbox2=\hbox{$#1\vcenter{}$}
        \setbox4=\hbox{\raisebox{-\ht2}[.2pt][.2pt]{$#1-$}}
        \hbox{$#1\blacktriangleleft$}
        \nointerlineskip
        \kern\wd0
        \copy4
    }
}
\makeatother
\usepackage{dsfont}

\usepackage{multirow}
\usepackage{kotex} 

\usepackage{mathtools}

\definecolor{my_blue}{RGB}{0,112,192}

\usepackage{placeins}

\usepackage{tikz}

\usepackage{fvextra}

\DefineVerbatimEnvironment{WideMinted}{Verbatim}
{breaklines, fontsize=\small, frame=lines, breakanywhere, xleftmargin=0pt, xrightmargin=0pt}

\usepackage{enumitem}
\usepackage{listings}
\definecolor{my_green}{rgb}{0.1, 0.5, 0.2}
\definecolor{my_pink}{rgb}{0.9, 0.3, 0.7}
\definecolor{my_blue}{rgb}{0.3, 0.5, 0.8}
\definecolor{my_brown}{HTML}{733120}
\definecolor{my_correct}{HTML}{91cb8c}

\usepackage{tablefootnote}

%% file: content/0_abstract.tex
\begin{abstract}
This work addresses test output prediction, a key challenge in test case generation. 
To improve the reliability of predicted outputs by LLMs, 
prior approaches generate code first to ground predictions. One grounding strategy is direct execution of generated code, but even minor errors can cause failures. To address this, we introduce \textit{LLM-based pseudocode execution}, which grounds prediction on more error-resilient pseudocode and simulates execution via LLM reasoning. We further propose \ours, a dual-execution framework that combines both approaches by \edithj{functional majority voting}. Our analysis shows the two approaches are complementary in overcoming the limitations of
direct execution suffering from code errors, and pseudocode reasoning from hallucination. 
On LiveCodeBench, \ours achieves the state-of-the-art performance, improving Pass@$1$ by 13.6 pp.
For filtering candidates in code generation, \ours shows the best Pass@$1$ on LiveCodeBench-Easy, BigCodeBench-Hard, DevEval and HumanEval(+).\footnote{\ourslink}
\end{abstract}

%% file: content/1_intro.tex
\section{Introduction}
\label{sec:intro}
Test cases play a critical role in code generation, serving as tools to evaluate and refine model outputs~\cite{chen2023codet,shinn2023reflexion,he-etal-2024-cocost,han-etal-2024-archcode,huang-etal-2024-enhancing}.
For instance, integrating test case generation into the code generation pipeline enables automatic evaluation, allowing LLMs to generate multiple code candidates and filter them based on their predicted test outputs~\cite{chen2023codet,han-etal-2024-archcode}. 

\begin{figure}[t]
    \centering
    \begin{tabular}{@{}c@{}}
        \hspace{-9px}
        \subfloat[Direct Code Execution (TestChain). \label{fig:1a}]
        {\includegraphics[width=0.75\linewidth]{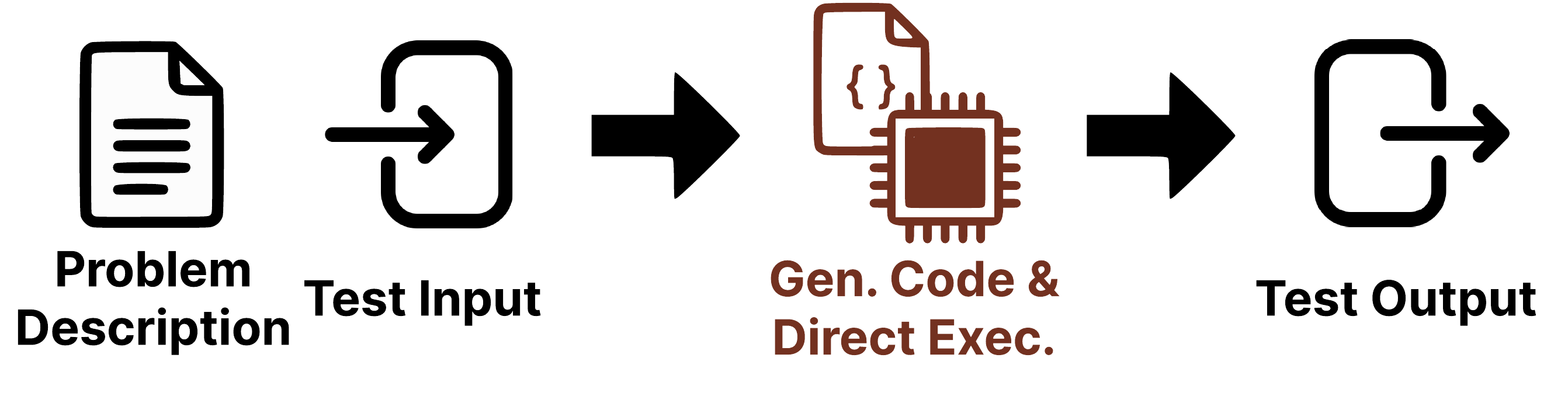}} \\
        \hspace{-9px}
        \subfloat[LLM-based Pseudocode Execution.
        \label{fig:1b}]
        {\includegraphics[width=0.75\linewidth]{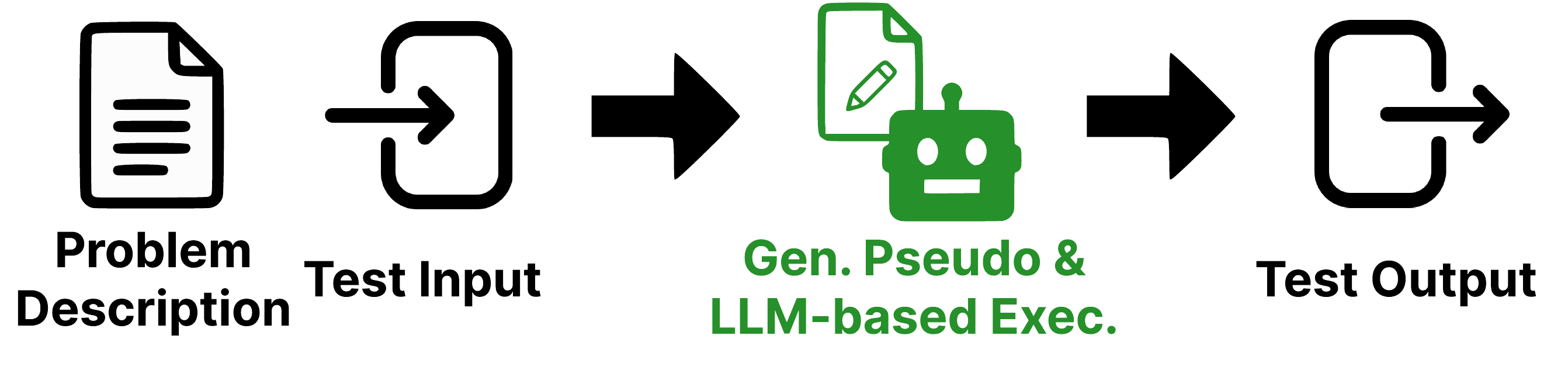}} \\[-7pt]
        \hspace{-9px}
        \subfloat[Dual Execution (\ours). 
        \label{fig:1c}]
        {\includegraphics[width=0.75\linewidth]{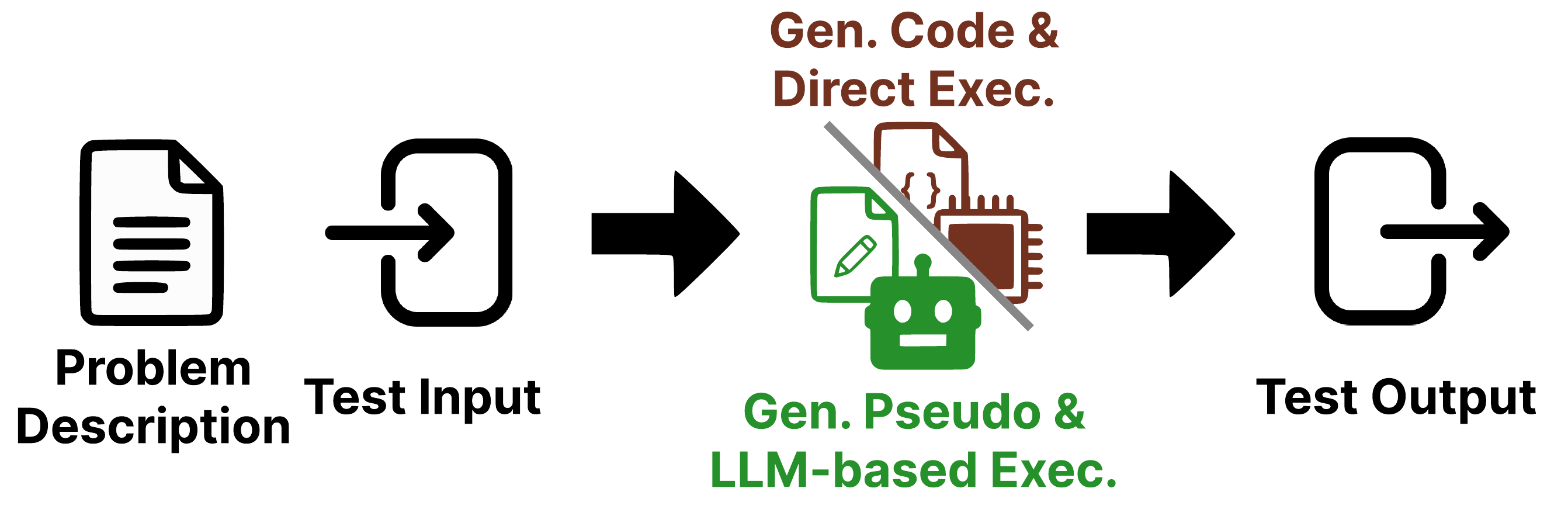}} \\
    \end{tabular} 
    \caption{
    Comparison of test output prediction methods:
    (a) \textbf{TestChain}~\cite{li2024largelanguagemodelstest} grounds generated code and executes it directly.  
    (b) \textbf{LLM-based Pseudocode Execution} grounds generated pseudocode and simulates its execution via an LLM.  
    (c) \textbf{\ours} combines both code and pseudocode grounding, executing them via direct and LLM-based paths, respectively.
}
    
    \label{fig:overview}
\end{figure}

\begin{figure*}[t]
    \centering
    \begin{tabular}{@{}cc@{}}
        \subfloat[
            Direct execution failure.\label{fig:2a}]{
            \includegraphics[height=0.4\linewidth]{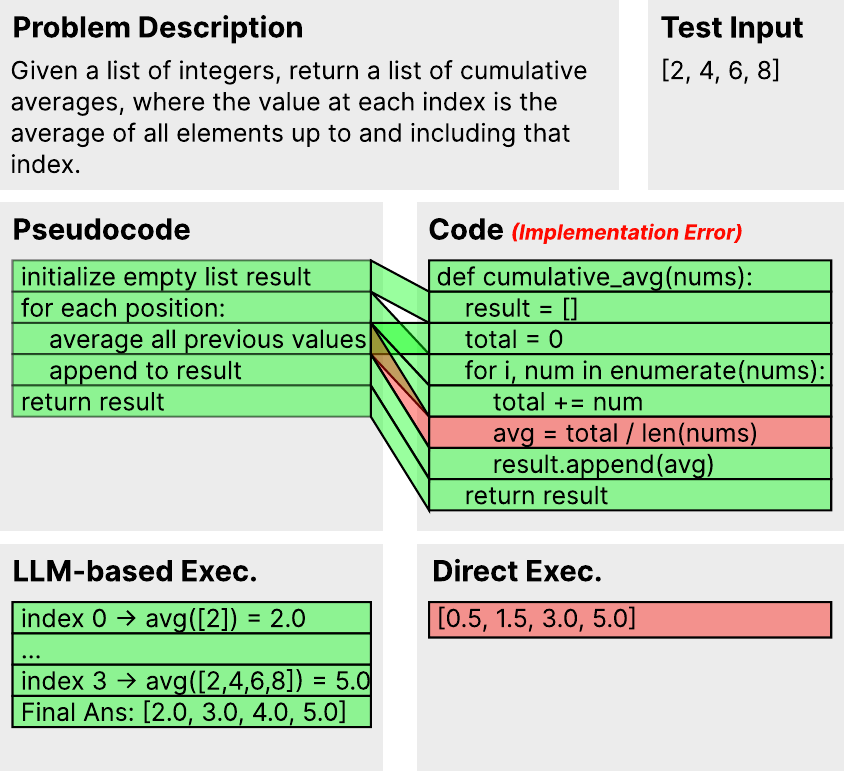}
        } &
        \subfloat[
            Pseudocode execution failure.\label{fig:2b}]{
            \includegraphics[height=0.4\linewidth]{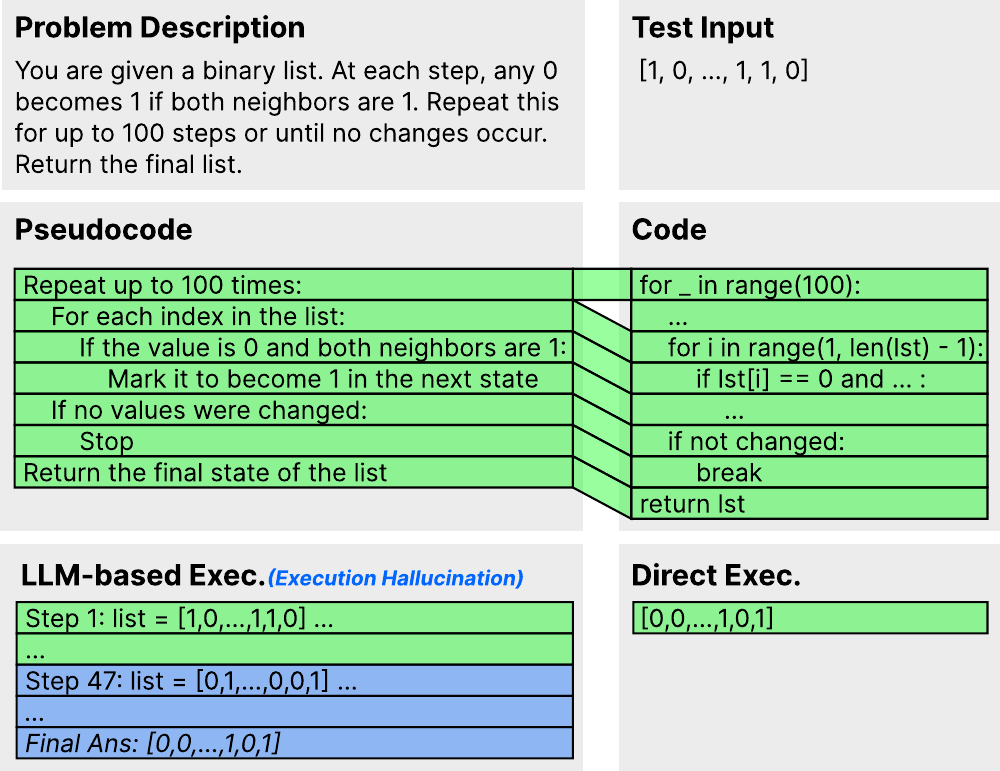}
        }   \\
    \end{tabular}
    \caption{
    (a) The problem involves complex functional logic that is accurately captured in the generated pseudocode but {\color{red!50} mistranslated into executable code (i.e., implementation errors)}.
    (b) Both the generated code and pseudocode are correct, but simulating the pseudocode (e.g., involving deeply nested loops) introduces {\color{blue!50}reasoning challenges that lead to LLM prediction errors (i.e., execution hallucinations)}.
    }
    \label{fig:text_example}
\end{figure*}

Test case generation can be categorized into input generation and output prediction (\citealp{li2024largelanguagemodelstest}).
In this work, we mainly focus on the output prediction, which requires precise program reasoning~\cite{NEURIPS2023_58168e8a,liu2023evaluatinglogicalreasoningability,brown2024largelanguagemonkeysscaling}, 
considered difficult despite recent LLM advances~\cite{barr2014oracle,li2024largelanguagemodelstest}.\footnote{{We note that test output prediction is orthogonal to input generation and valuable in its own right: for instance, while AlphaCode~\cite{li2022competition} focuses solely on test input generation, \textsc{CodeT}~\cite{chen2023codet} additionally predicts test outputs and achieves stronger empirical performance on end-to-end code generation.}}
Recent studies like TestChain~\cite{li2024largelanguagemodelstest} improve prediction performance by grounding it on generated code (Figure~\ref{fig:1a}); {however, their reliability remains brittle, as models may produce the correct logic (e.g., correct pseudocode) but fail to realize it due to minor implementation errors~(\citealp{zhong-etal-2020-semantic-scaffolds,jiang2024self}; Figure~\ref{fig:2a})}.

Our first contribution is to deconfound correct logic from implementation errors in generated code.
We propose \textit{LLM-based pseudocode execution} that employs  pseudocode in place of generated code then simulates execution by LLM reasoning (Figure~\ref{fig:1b}; \S\ref{sec:proposed_base}).
By expressing the algorithmic intent at a higher level of abstraction, pseudocode 
bypasses minor implementation
errors.
While this approach offsets the weaknesses of direct execution, LLM-based execution may generate hallucinated reasoning steps (an execution hallucination) that diverge from the intended logic (\citealp{kojima2022large,luo2024improve}; Figure~\ref{fig:2b}).

Our second contribution is to propose \ours that integrates direct code execution and LLM-based pseudocode execution using \edithj{functional majority voting~\cite{li2022competition,chen2023codet,launer2026majority}} (Figure~\ref{fig:1c}; \S\ref{sec:proposed_main}). 
Figure~\ref{fig:ours} illustrates why the two approaches are effective in combination:
(1) For problems
that are susceptible to  implementation errors (Figure~\ref{fig:2a}), pseudocode offers a more reliable grounding by capturing the algorithmic intent without being constrained by syntactic or low-level details (\S\ref{sec:analysis_code_correctness}).
(2) When execution hallucinations are more likely (Figure~\ref{fig:2b}), direct execution offers more dependable results due to its deterministic nature. It computes the output exactly as specified, without relying on LLM inference (\S\ref{sec:analysis_reasoning}).

We evaluate \ours
on the LiveCodeBench test output prediction benchmark (\citealp{LiveCodeBench}; \S\ref{sec:test_output_prediction}).
\ours achieves 13.6 pp gain over the previous state-of-the art and 5.6 pp gain over TestChain of using direct code execution only.

For end-to-end code generation (\S\ref{sec:code_generation}), integrating \ours into \textsc{CodeT}~\cite{chen2023codet} improves the Pass@$1$ code generation performance of \texttt{Llama-3.1-8B-Instruct}~\cite{grattafiori2024llama} by 3.2 pp. 
Notably, replacing it with TestChain hurts the performance by 5.6 pp. 
This is because direct code execution suffers from the zero-advantage problem (\citealp{yu2025dapo,le2026no}; \S\ref{appendix:why_testchain_bad}): it can yield incorrect outputs when the generated code is flawed, reducing its effectiveness for filtering candidate programs based on predicted outputs. In contrast, LLM-based pseudocode execution is decoupled from code correctness and thus more robust in this setting.


Our key contributions are as follows:
\begin{itemize}
    \item \edit{\textbf{First use of LLM-based pseudocode execution} for test-output prediction.}
    \item \edit{\textbf{A novel dual-path framework (\ours)} that leverages the complementary strengths of direct and pseudocode execution paths.}
    \item \edit{\textbf{State-of-the-art} performance in \textbf{test output prediction} on LiveCodeBench and \textbf{best} performance in \textbf{end-to-end code generation} across LiveCodeBench-Easy, BigCodeBench-Hard~\cite{zhuo2024bigcodebench}, and HumanEval(+)~\cite{codex,YourCodeGenerated}.} 
    \item \edit{\textbf{Identification of the zero-advantage problem} in end-to-end code generation.}
\end{itemize}


%% file: content/2_preliminary_hj.tex
\section{Preliminary and Related Work}
\label{sec:preliminary}
In this section, we present the formal definition of test output prediction and its standard solution via direct code execution. {Due to space limit, we leave an extensive survey in Appendix~\ref{sec:related}.}

\subsection{Problem: Test Output Prediction}
\label{sec:prelim:task}
\paragraph{Test Case Generation}
Test case generation involves two stages:
(1) input generation, which produces a valid test input based on the problem description; and
(2) output prediction, which infers the corresponding output given the input and the problem.



While input generation has progressed through LLMs~\cite{li2022competition} and fuzzing~\cite{deng2023large,xia2024fuzz4all}, output prediction remains difficult~\cite{barr2014oracle,li2024largelanguagemodelstest}. 

\paragraph{Test Output Prediction}
Formally, let $\mathcal{D}$, $\mathcal{I}$, and $\mathcal{O}$ denote the spaces of problem descriptions, test inputs, and outputs, respectively.
We define output prediction as the function ${f}: \mathcal{D} \times \mathcal{I} \rightarrow \mathcal{O}$, which maps a problem description and test input to the corresponding output.

\subsection{Baseline: Direct Code Execution}
\label{sec:prelim:code_exec}
When correct code is available, test output prediction $f$ reduces to a simple code execution function $e: \mathcal{C} \times \mathcal{I} \rightarrow \mathcal{O}$ that runs the code on the input and returns the result.
The set $\mathcal{C}$ denotes the space of code implementations.

The goal of \textit{direct code execution} baseline like TestChain~\cite{li2024largelanguagemodelstest} is to synthesize code that approximates the ground-truth implementation, such that its execution predicts test outcome.

Test output prediction $f$ can thus be formulated via a code generator $g: \mathcal{D} \rightarrow \mathcal{C}$, followed by code execution $e$:
\begin{align}
  {f}(d, i) = {e}({g}(d), i),
\end{align} where $d \in \mathcal{D}$ is the problem description and $i \in \mathcal{I}$ denotes the test input.


%% file: content/3_proposed_hj.tex
\section{Proposed Method}
\label{sec:proposed}
To improve test output prediction, we first deconfound direct execution baseline with pseudocode (\S\ref{sec:proposed_base}).
We then propose \ours (\S\ref{sec:proposed_main}), a dual-execution strategy that combines direct and LLM-based execution via {\textit{path-weighted functional majority voting}, which accounts for intra-path agreement and adjusts voting weights accordingly}.

\subsection{Our Distinction:  Pseudocode Grounding}
\label{sec:our_distinction}
Direct execution for test output prediction relies on a strong assumption that \textit{the generated code is correct}. {However, as illustrated in Figure~\ref{fig:2a}, even minor implementation errors introduced by the generator can significantly undermine prediction reliability~\cite{zhong-etal-2020-semantic-scaffolds,jiang2024self}.}

{Our design choice is to avoid relying on generated code, as prediction correctness becomes entangled with implementation quality. Instead, we ground test output prediction on pseudocode, which abstracts away low-level implementation details {(see Appendix~\ref{appendix:prompt_engineering} for details)}.}

Formally, we disentangle code generation $g$ into two modules:
$g = t \circ p$, where $p: \mathcal{D} \rightarrow \mathcal{P}$ generates high-level pseudocode and $t: \mathcal{P} \rightarrow \mathcal{C}$ translates it into executable code~\cite{kulal2019spoc, huang2023codecot, jiang2024self, islam2024mapcoder}.\footnote{\edit{Prior work has shown that introducing this additional pseudocode step improves code generation quality compared to direct generation from descriptions~\cite{jiang2024self}.}}

Unlike direct execution using $g(d) = t \circ p(d)$, we bypass the translation step $t$ and ground prediction directly on pseudocode $p(d)$.
By doing so, we avoid errors introduced during translation, such as syntactic inaccuracies, execution failures, or misindexed operations, which are a major source of unreliability in $g$.


\begin{figure}[t]
    \centering
    \includegraphics[width=1.0\linewidth]{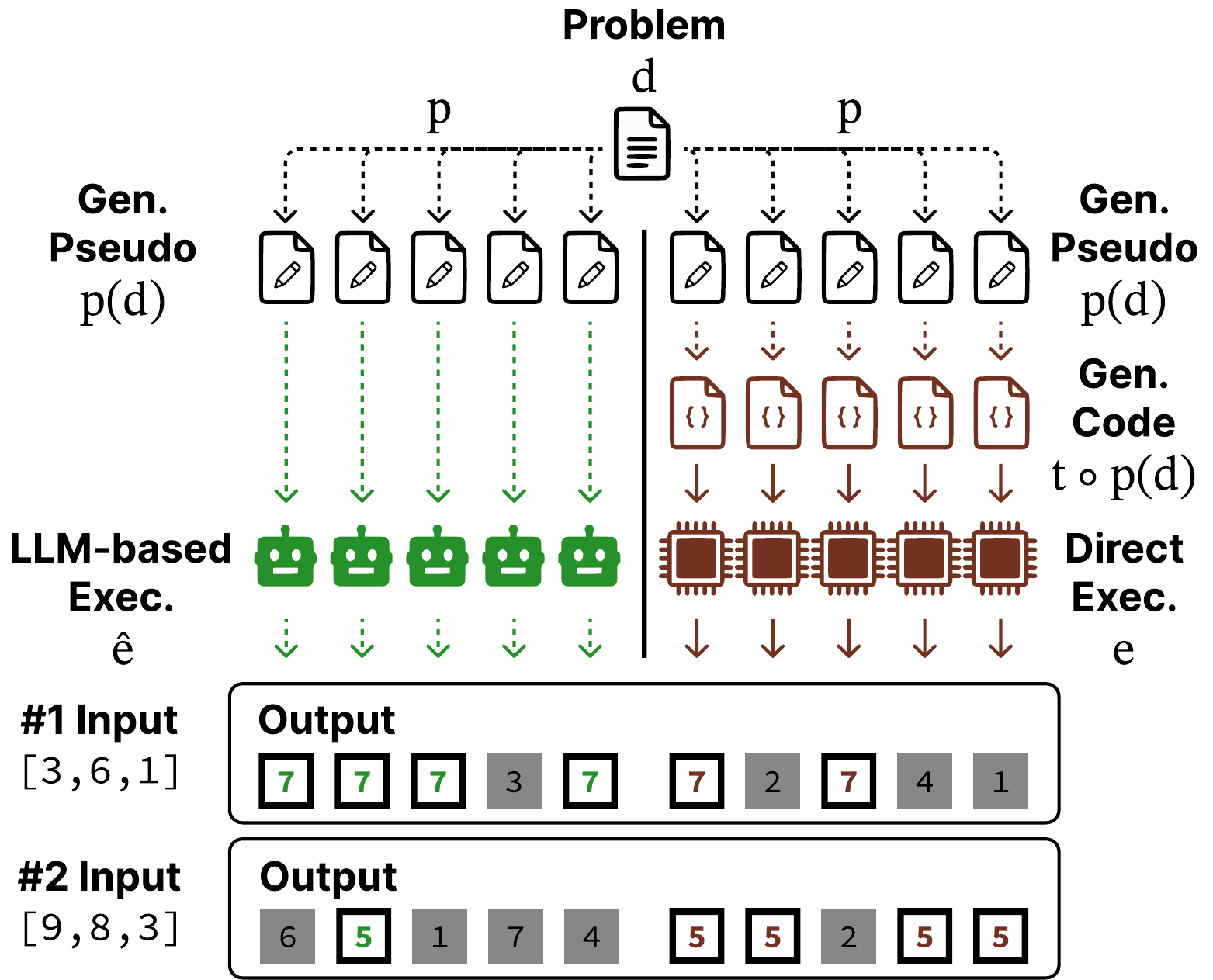}
    \caption{
        Overview of our dual-execution framework for test output prediction.
        Given a problem description and input, we perform two separate executions in parallel.
        In the left path, we generate pseudocode and use an LLM to reason over it to predict the output (i.e., {\textcolor{my_green}{LLM-based pseudocode execution}}).
        In the right path, we generate pseudocode independently, translate it into executable code, and run it on the input to obtain the output (i.e., {\textcolor{my_brown}{direct code execution}}).
        Multiple outputs from both paths are aggregated via \edithj{path-weighted functional majority voting} to determine the final prediction.
    }
    \label{fig:ours}
\end{figure}

\subsection{\ours}
\subsubsection{LLM-based Pseudocode Execution}
\label{sec:proposed_base}
Since pseudocode is not directly executable, we simulate its behavior using a language model, denoted as \textit{LLM-based execution}.
This yields the prediction function:
\begin{align}
    \hat{f}(d, i) = \hat{e}(d, p(d), i),
\end{align} where $\hat{e}$ reasons over the pseudocode $p(d)$ and input $i$ to produce an output.

{While LLM-based pseudocode execution avoids implementation errors by not relying on executable code, it remains susceptible to \textit{execution hallucinations}, i.e., failures that arise when the model produces plausible but incorrect intermediate reasoning steps that deviate from the intended logic (\citealp{kojima2022large,luo2024improve}; Figure~\ref{fig:2b}).
These are distinct from failures in direct code execution, where the model does not simulate execution through reasoning but instead directly runs the generated code. Such failures reflect implementation errors in the code itself.}


\subsubsection{Dual Execution}
\label{sec:proposed_main}
To overcome the limitations of relying on a single execution path, we propose \ours, a dual-execution framework that combines direct code execution and LLM-based pseudocode execution. 

Our key contribution is to identify that these two approaches succeed under different conditions.
Direct execution guarantees correctness when the generated code is correct, but suffers from brittle failure modes due to syntactic or semantic bugs (e.g., missing edge cases, off-by-one errors).
Conversely, pseudocode execution is more resilient to such implementation issues by relying on high-level algorithmic intent, but can struggle when the reasoning chain is long or complex.
As illustrated in Figure~\ref{fig:text_example}, pseudocode execution can outperform code execution when the generated code is difficult to synthesize correctly, while code execution becomes preferable when the reasoning burden becomes too high for LLMs. 

As shown in Figure~\ref{fig:ours}, we thus combine both execution methods and aggregate their outputs using \edithj{functional majority voting (FMV; ~\citealp{li2022competition,chen2023codet,launer2026majority})}.
This is well-suited for integrating heterogeneous results from both paths (see Table~\ref{tab:combination}).
This straightforward formulation also makes it easy to extend (see Appendix~\ref{appendix:method_simplicity} for details).
{To further reflect confidence in each execution route, we propose a \edithj{\textit{path-weighted} variant of FMV} that boosts the influence of outputs from routes with unanimous agreement during final aggregation.}

Formally, given a problem description $d \in \mathcal{D}$ and a test input $i \in \mathcal{I}$, we collect $l$ outputs from direct execution and $m$ outputs from LLM-based pseudocode execution. 
Although we set $l{=}m$ in our experiments for simplicity, the framework allows them to differ in practice depending on resource allocation {(see Appendix~\ref{appendix:overhead} for computational overhead analysis)} or confidence in each execution method.

\paragraph{\edithj{\textbf{Path-Weighted FMV}}}
\edithj{We extend FMV by introducing path-wise confidence weighting.
Each path casts votes proportional to its matching outputs. If all valid outputs within a path unanimously agree on the same result, its votes receive a binary boost to \(w_{high}\) (otherwise \(w_{base}\)). Formally, let $O$ and $\hat{O}$ be multisets of outputs from the direct ($f$) and LLM-based ($\hat{f}$) execution paths. The final output is selected as follows:}
\edithj{\begin{align}
    \label{eq:final_output}
    o &=  \arg\max_{o \in O \cup \hat{O}} [w(o, O) \cdot V(o,O) \nonumber \\
       &\phantom{MMMMl} + w(o, \hat{O}) \cdot V(o,\hat{O})], \nonumber \\
    w(o,O) &= \begin{dcases}
                   w_{high},  \textrm{ if } \forall o_i \in O, o_i=o,\\
                   w_{base},            \textrm{ otherwise,}
                \end{dcases} \nonumber\\ 
    V(o,O) &= |\{i : o_i \in O, o_i=o\}|,
\end{align}}
where $w_{high} > w_{base}$.

{This strategy introduces no additional cost and consistently yields slight performance gains without degradation, as shown in Appendix~\ref{appendix:consensus}.}

%% file: content/4_experiments.tex
\section{Experimental Setup}
\label{sec:exp}

\subsection{Benchmark}
\paragraph{Test Output Prediction}
We evaluate our method on LiveCodeBench~\cite{LiveCodeBench}, a benchmark covering four code-related tasks: test output prediction, code execution,\footnote{\edit{Note that test output prediction differs from the code execution task, which is the primary focus of benchmarks like CRUXEval~\cite{pmlr-v235-gu24c}, as detailed in Appendix~\ref{appendix:cruxeval}.}} code generation, and self-repair.
Our main focus is on the test output prediction task, which we evaluate under two knowledge cutoff settings to prevent data contamination: (i) \textbf{May 1, 2023–Apr 1, 2024 (442 problems)}: Larger and more representative, but some models may have exposure to the data; (ii) \textbf{Jan 1–Apr 1, 2024 (76 problems)}: Enables comparison across multiple models with minimal contamination risk.

\input{table/leaderboard}

\paragraph{Code Generation}

\edit{To apply our method to post-generation filtering (\S\ref{sec:code_generation}), we evaluate code generation on LiveCodeBench-Easy (554 problems, Jan 2024–Feb 2025) for competitive programming, and three realistic non-OJ scenarios: tool-integrated BigCodeBench-Hard~\cite{zhuo2024bigcodebench} (\S\ref{analysis:bigcodebench}), repo-level DevEval~\cite{li-etal-2024-deveval} (\S\ref{analysis:repo}), and HumanEval(+) (Appendix~\ref{appendix:end2end_humanevalplus}).}

\subsection{Metric}
\edit{We evaluate performance using Pass@$k$~\cite{codex}, adopting the \textit{ranking-based formulation}~\cite{chen2023codet} where the model selects the top-$k$ candidates from generated samples.
Regarding sample size, we set $n{=}10$ for baselines and $l{=}m{=}10$ in \ours for test output prediction, while setting $l{=}m{=}5$ per test input for end-to-end code generation.}



\subsection{Test Output Prediction Baselines}
{For all implementation details, refer to Appendix~\ref{subsec:implementation}.}
 
\paragraph{No Grounding} The model predicts test outputs directly from the problem description and input, without grounding in any intermediate representations.
This setting can be regarded as a variant of \textsc{CodeT}~\cite{chen2023codet} that uses test output prediction to rank candidate programs, isolating its output prediction component from the full pipeline.
\paragraph{TestChain\textrm{~\cite{li2024largelanguagemodelstest}}} A grounded test output prediction baseline that predicts outputs by generating code from the input and executing it to obtain the final answer.



\section{Experimental Results}
\label{sec:results}

\input{table/cutoff}

\input{table/code_filtering.tex}

\subsection{Test Output Prediction}
\label{sec:test_output_prediction}
\paragraph{Main Leaderboard Results (Table~\ref{tab:leaderboard})}
We report Pass@$1$ results on the full LiveCodeBench test output prediction benchmark (May 1, 2023–April 1, 2024).
Among uncontaminated models, \texttt{GPT-4-Turbo-2024-04-09} served as the strongest baseline.
\edit{With this model as the backbone,} \ours achieves a new state-of-the-art Pass@$1$ of 81.1, outperforming \texttt{GPT-4-Turbo} with FMV by 6.9 pp and TestChain with FMV by 5.6 pp.
\edit{This result is consistent with models potentially suffering from data contamination (see Appendix~\ref{appendix:exaone}).} 
These results confirm the benefit of combining both execution modes.

\paragraph{Expanded Comparison with Broader Model Set (Table~\ref{tab:cutoff})}
To compare methods across a wider range of LLMs, we apply them to a contamination-free subset of LiveCodeBench (January 1–April 1, 2024).
\ours achieves the highest Pass@$1$ among grounded methods across all LLM backbones, 
confirming the benefit of combining code and pseudocode execution.\footnote{{We further analyze test output prediction results by problem and input difficulty in Appendix~\ref{appendix:input_difficulty}.}}

\begin{figure}[t]
    \centering
    \includegraphics[width=1.0\linewidth]{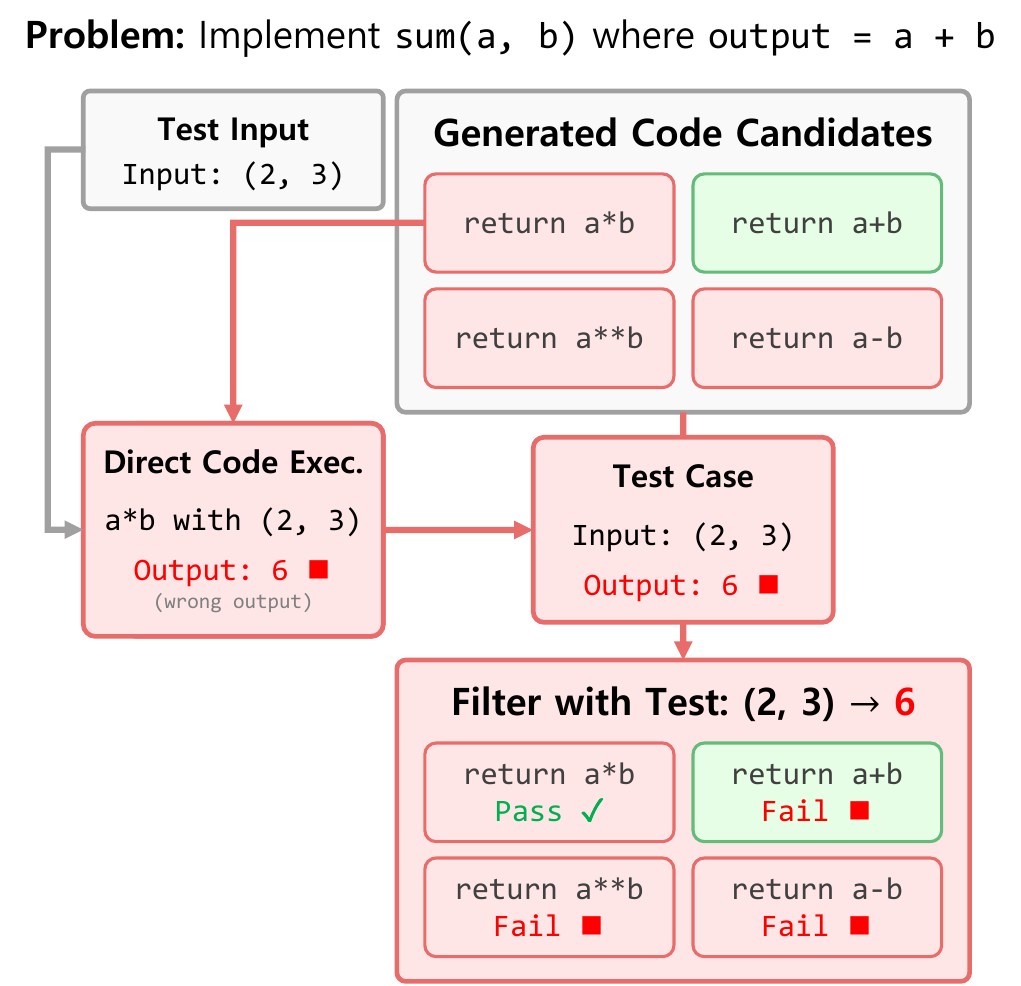}
    \caption{
    {Illustrative example of TestChain's zero-advantage problem in end-to-end code generation, \edit{where test input (\texttt{Input: (2, 3)}) is generated by \textsc{CodeT}.} }
    }
    \label{fig:circular_dependency}
\end{figure}
\subsection{Impact on Code Generation}
\label{sec:code_generation}

We evaluate whether improved test output prediction transfers into better code generation when applied to the code filtering stage~\cite{chen2023codet,han-etal-2024-archcode}. 
For each problem, five test inputs are generated using \textsc{CodeT}~\cite{chen2023codet} and 10 output predictions per input are aggregated via path-weighted FMV. Then the generated 5 test input output pairs are used to filter \edit{20} candidate code snippets.



Table~\ref{tab:code_filtering} shows that performance depends on how test outputs are obtained.  TestChain performs 5.6~pp worse than \textsc{CodeT}, possibly due to the zero-advantage problem in direct code execution (\S\ref{appendix:why_testchain_bad}; Appendix~\ref{appendix:end2end_humanevalplus}).
In contrast, LLM-based pseudocode execution is free from such problem and 
improves over \textsc{CodeT} by 1.7 pp in Pass@$1$ for end-to-end code generation. 
\ours achieves the best performance for all Pass@$k$ metrics.

\section{Analysis}
\label{sec:analysis}


\subsection{Zero-Advantage Problem in TestChain}
\label{appendix:why_testchain_bad}
As shown in Table~\ref{tab:code_filtering}, TestChain (direct code execution only) hurts final Pass@$1$ in end-to-end code generation. We hypothesize that inferring test outputs from the candidate programs themselves and then reusing them to rank the same pool is structurally limited (Figure~\ref{fig:circular_dependency}). 
Analogous to the zero-advantage problem in GRPO-style RL~\cite{yu2025dapo,le2026no}, the induced outputs provide little signal when most candidates are correct and become unreliable when most are wrong. As a result, filtering is least informative when it is most needed. \ours mitigates this issue by introducing an orthogonal pseudocode-based execution path whose predictions are not directly tied to the candidate code being ranked, consistent with the gains in Table~\ref{tab:code_filtering}.


\begin{figure}[t]
    \centering
    \includegraphics[width=1.0\linewidth]{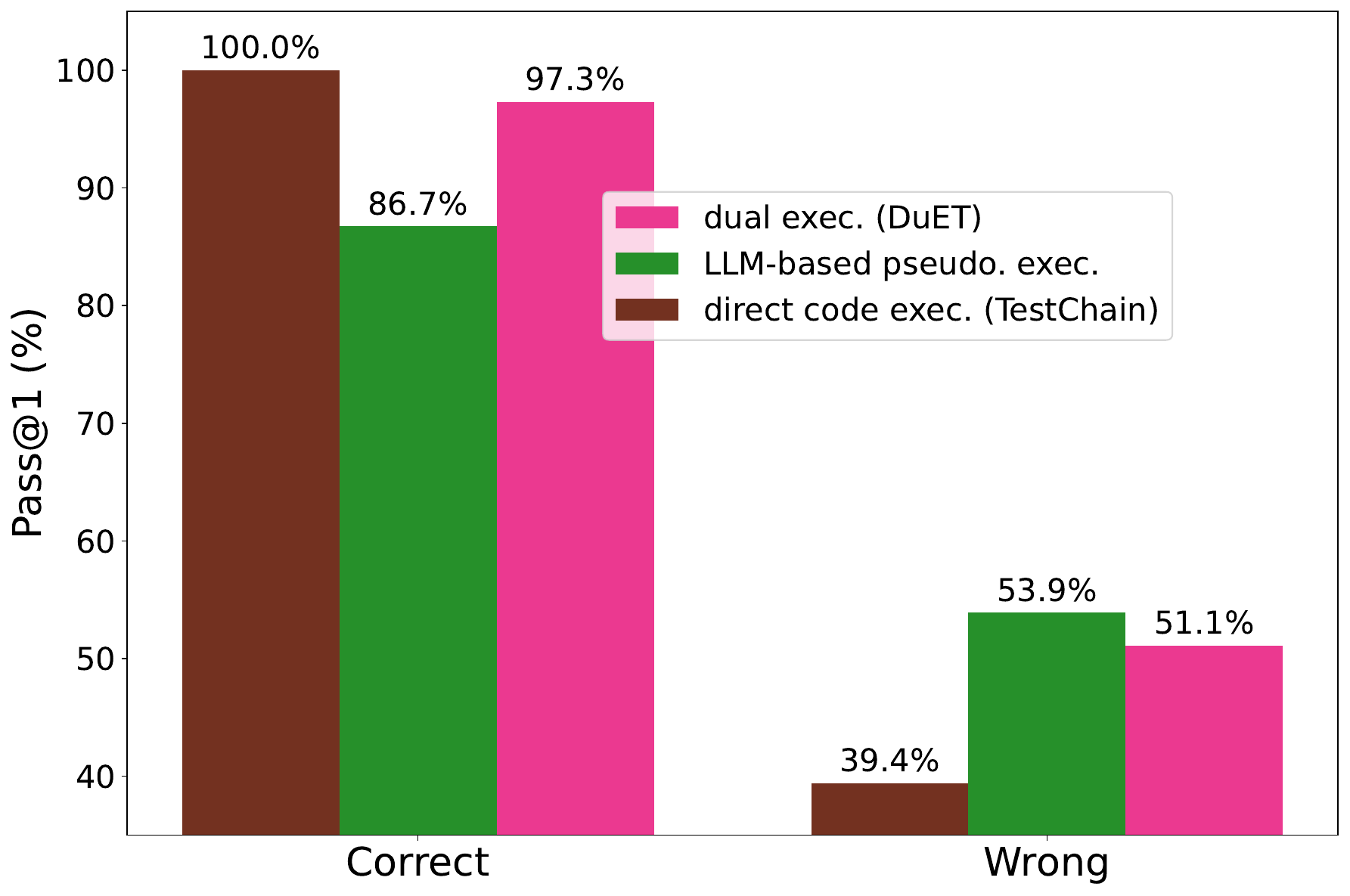}
    \caption{
        Pass@$1$ scores of execution methods on LiveCodeBench test output prediction (Jan 1–Apr 1, 2024) by \texttt{GPT-4-Turbo-2024-04-09}, across problem subsets based on the generated code correctness.
    }
    \label{fig:code_correctness_bar}
\end{figure}

\subsection{Impact of Generated Code Correctness}
\label{sec:analysis_code_correctness}

Execution accuracy is highly sensitive to implementation errors in generated code.
To analyze this effect, we group problems in Figure~\ref{fig:code_correctness_bar} based on whether the generated code passes all test inputs (\textit{Correct}) or fails at least one (\textit{Wrong}).

In the \textit{Correct} group, \textcolor{my_brown}{direct code execution} naturally achieves 100, while \textcolor{my_green}{LLM-based pseudocode execution} reaches 86.7, limited by occasional execution hallucinations.
\textcolor{my_pink}{\ours} achieves 97.3,
primarily by leveraging correct outputs from direct execution while mitigating hallucinations through path-weighted FMV.

In the \textit{Wrong} group, \textcolor{my_brown}{direct code execution}
drops sharply to 39.4 due to implementation errors, while
\textcolor{my_green}{LLM-based pseudocode execution} is more robust at 53.9.
\textcolor{my_pink}{\ours} achieves 51.1, generally benefiting from pseudocode over faulty code.


These results suggest that \ours benefits from pseudocode execution when generated code is faulty, and from direct execution when code is correct.

\input{table/design_space}

\subsection{{Robustness Across Execution Trace Lengths}}
\label{sec:analysis_reasoning}
Execution trace length, defined as the number of execution steps~\cite{liu-etal-2023-code} taken by generated code, reflects the complexity of reasoning required to compute outputs.
To assess its impact on prediction accuracy, we group samples by trace length and measure Pass@$1$ for each group.


\begin{figure}[t]
    \centering
    \includegraphics[width=1.0\linewidth]{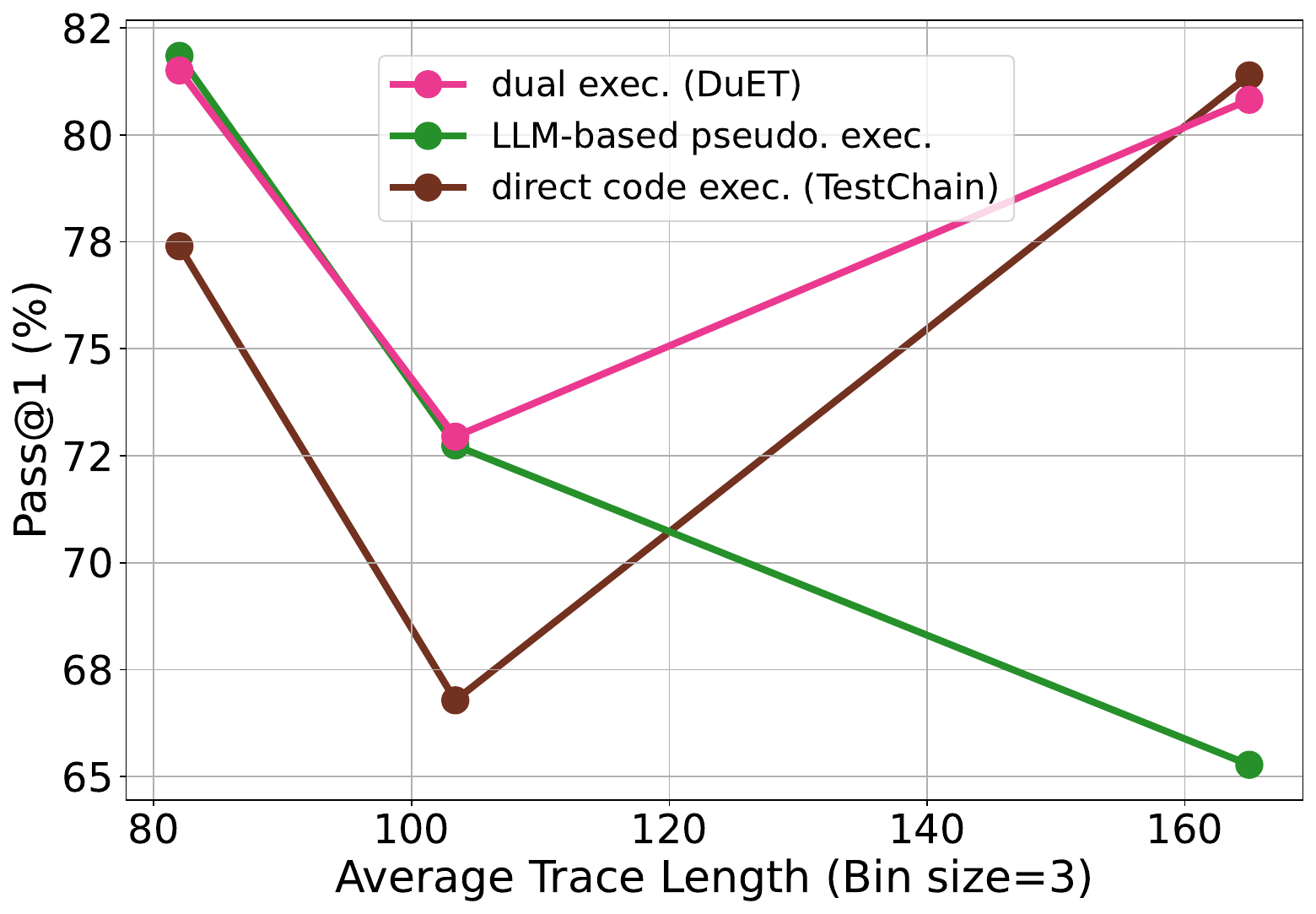}
    \caption{
        Pass@1 across bins of generated code execution trace lengths on LiveCodeBench test output prediction (Jan 1–Apr 1, 2024) by \texttt{GPT-4-Turbo-2024-04-09}.
        Each execution step corresponds to a single line of code executed during code execution.
    }
    \label{fig:trace_length_histogram}
\end{figure}

\input{table/combination}

{As shown in Figure~\ref{fig:trace_length_histogram}, we observe a clear trend: Across \textbf{all trace lengths}, \ours of {\textcolor{my_pink}{dual execution}} shows consistently strong performance, {matching or surpassing} competing methods. 
{\ours of hybrid execution is specifically designed to mitigate errors in both paths:} up to around 100 trace steps, \textcolor{my_green}{LLM-based pseudocode execution} often outperforms direct code execution by leveraging the LLM's reasoning capabilities. For longer traces, however, its performance degrades as multi-step reasoning becomes error-prone. Here, \textcolor{my_brown}{direct execution} proves more reliable as it is immune to such brittleness.}

{We note that the decline in LLM-based execution with longer traces does not indicate a fundamental limitation of the LLM itself. Rather, it reflects a natural characteristic of multi-step reasoning: as the number of steps increases, even a single mistake can derail the entire computation, making long sequences inherently more error-prone~\cite{wu2025more,patel2024multi}. Nonetheless, LLM-based pseudocode execution remains effective for traces of up to $\sim$100 steps, while the full \ours system combines both execution modes to adapt to trace complexity and maintain high accuracy.}

\subsection{Design Space for Grounded Prediction}
\label{sec:llm_based_code_exec}
We explore the design space of grounded test output prediction (summarized in Table~\ref{tab:design_space}), defined by two orthogonal choices: the execution target (code vs. pseudocode) and the execution mode (direct vs. LLM-based). 
Table~\ref{tab:combination} reports Pass@$1$ results for all three feasible combinations, allowing us to compare the relative strengths of each strategy.


\input{table/bigcodebench_hard_qwq}

\paragraph{Code: {Direct vs. LLM-based Execution}}
Direct code execution outperforms LLM-based code execution (\textcolor{my_brown}{75.5} vs. \textcolor{my_blue}{72.2}) by faithfully reflecting actual runtime behavior, whereas the latter remains prone to reasoning failures.
Consistently, direct execution achieves superior performance over LLM-based code execution when combined with LLM-based pseudocode execution (\textcolor{my_pink}{80.8} vs. \textcolor{my_blue}{73.6}). 

\paragraph{LLM-based Execution: {Pseudocode vs. Code}}
Pseudocode offers a more reliable grounding than code for LLM-based execution (\textcolor{my_green}{74.1} vs. \textcolor{my_blue}{72.2}), as it enables the model to focus on high-level logic.
When paired with direct execution, pseudocode again yields better performance than code in LLM-based execution, helping reduce implementation errors (\textcolor{my_pink}{80.8} vs. \textcolor{my_blue}{77.4}).


These results support the combination of direct code execution and LLM-based pseudocode execution for the best performance.


\subsection{{On Realistic Benchmark with Reasoning-Enhanced LLM}}
\label{analysis:bigcodebench}
{We further conducted experiments with the reasoning model QwQ-32B~\cite{qwq32b} on the more challenging  BigCodeBench-Hard, \edit{which involves diverse function calls as tools for tasks such as data analysis and web development.}
In Table~\ref{tab:qwq32b_bigcodebench}, 
\ours is \edit{{the only approach} that {outperforms \textit{No Filtering}}}, achieving the best performance (25.3 Pass@$1$).
This shows that \ours generalizes to reasoning-based LLMs and remains effective even when test outputs involve arbitrary external library objects.}

\begin{figure}[t]
    \centering
    \includegraphics[width=1.00\linewidth]{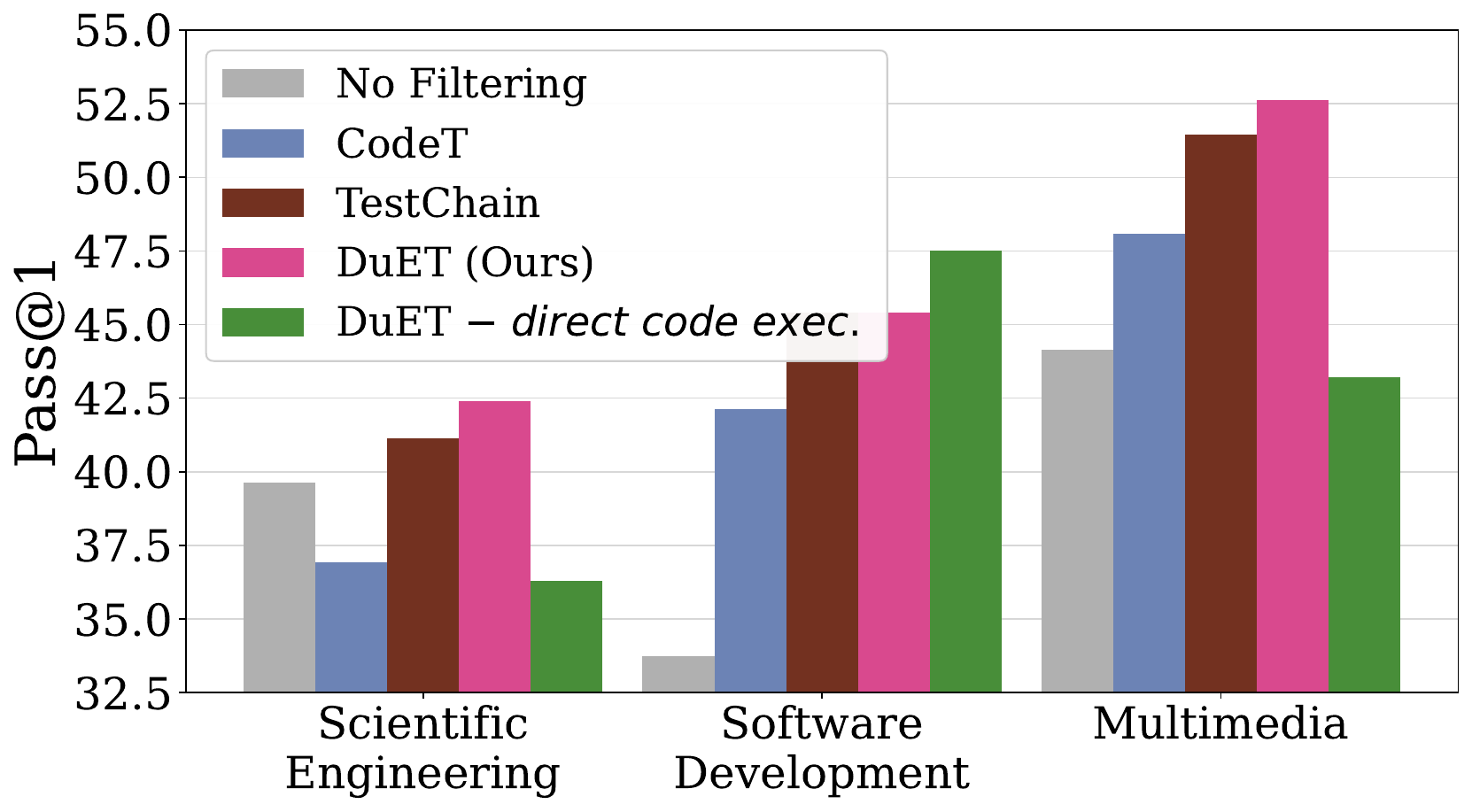}

    \caption{
        \edit{End-to-end code generation performance of \texttt{Llama-3.1-8B-Instruct} evaluated on DevEval (See Appendix~\ref{appendix:repo-level} for the full results).}
    }
    \label{fig:repo-level}
\end{figure}

\subsection{\edit{\textbf{Repo-level Code Generation}}}
\label{analysis:repo}
\edit{We further validated \ours on the repository-level benchmark, DevEval~\cite{li-etal-2024-deveval}; detailed experimental setups are provided in Appendix~\ref{appendix:repo-level}.
As shown in Figure~\ref{fig:repo-level}, the results align with the observations in Figure~\ref{fig:code_correctness_bar}: LLM-based pseudocode execution outperforms direct execution (TestChain) in challenging domains where baseline generation quality (\textit{No Filtering}) is low, whereas direct execution dominates in easier domains.
By combining both paths, \ours consistently achieves top-tier performance, ranking first or second across all domains.}

%% file: table/leaderboard.tex
\begin{table}[tb!]
    \centering
    \small
    \renewcommand{\arraystretch}{1.2}
    \begin{tabular}{lcccc}
        \thickhline
        \textbf{Method}                                                                            & \textbf{Pass@$1$}            \\
        \hline
        \small \texttt{Llama-3-8B-Instruct}\textsuperscript{\textdagger}              & \num{27.2}                 \\
        \small \texttt{Mixtral-8x22B-Instruct}\textsuperscript{\textdagger}           & \num{45.6}                 \\
        \small \texttt{Mistral-Large}\textsuperscript{\textdagger}                    & \num{48.6}                 \\
        \small \texttt{Gemini-Pro-1.5-May}\textsuperscript{\textdagger}               & \num{48.6}                 \\
        \small \texttt{\color{red!50}Dracarys-Llama-3.1-70B-Instruct}\textsuperscript{\textdagger}  & \num{49.1}                 \\
        \small \texttt{\color{red!50}Dracarys2-Llama-3.1-70B-Instruct}\textsuperscript{\textdagger} & \num{52.1}                 \\
        \small \texttt{GPT-4-0613}\textsuperscript{\textdagger}                       & \num{54.4}                 \\
        \small \texttt{Claude-3-Opus}\textsuperscript{\textdagger}                    & \num{57.8}                 \\
        \small \texttt{GPT-4-Turbo-1106}\textsuperscript{\textdagger}                 & \num{58.4}                 \\
        \small \texttt{\color{red!50}Dracarys-72B-Instruct}\textsuperscript{\textdagger}            & \num{58.9}                 \\
        \small \texttt{\color{red!50}Dracarys2-72B-Instruct}\textsuperscript{\textdagger}           & \num{59.6}                 \\
        \small \texttt{GPT-4-Turbo-2024-04-09}\textsuperscript{\textdagger}           & \textbf{\num{67.5}}                 \\
        \small \texttt{\color{red!50}GPT-4o-2024-05-13}\textsuperscript{\textdagger}                & \num{73.5}                 \\
        \hdashline
        
        \small \texttt{GPT-4-Turbo-2024-04-09}  & \color{black} {\num{65.78431372549016}}       \\
        \phantom{0}+ \textit{functional majority voting}      & \color{black} {\num{74.24585219}}    \\
        \small TestChain                                                            & \color{black} {\num{69.42621920563099}}          \\
        \phantom{0}+ \textit{functional majority voting}      & \color{black}{\num{75.45248869}}    \\
        \rowcolor{gray!20}
        \small \textbf{\ours}                                                         & \color{black}{\textbf{\num{81.07088989}}} \\
        \thickhline
    \end{tabular}
    \caption{
        Test output prediction results on LiveCodeBench (May 1, 2023–Apr 1, 2024). Red-colored models have a knowledge cutoff after this date, introducing the possibility of data contamination.
        Daggered ({}\textsuperscript{\textdagger}) results are sourced from the leader board.\tablefootnote{\href{https://livecodebench.github.io/leaderboard.html}{https://livecodebench.github.io/leaderboard.html}}\tablefootnote{\edit{We treat unbiased Pass@$k$ as random selection (see Appendix~\ref{appendix:metric_comparison}).}}
    }
    \label{tab:leaderboard}
\end{table}

%% file: table/cutoff.tex
\begin{table}[tb!]
    \begin{flushleft}
    \small
    \renewcommand{\arraystretch}{1.1}
    \resizebox{1.0\linewidth}{!}{
    \begin{tabular}{lll >{\color{black}}c ccc}
        \thickhline
        \textbf{Method}          & \textbf{Ground.} & \textbf{Exec.} & \color{black}{\textbf{Pass@$1$} }                                                                                                              \\
        \hline
        \small \texttt{GPT-4-Turbo-2024-04-09}   & -                                   & -                                   & {\num{72.3684210526315}}\\
        \small TestChain            & C                                & D                              & {\num{74.34210526}} \\
        
        \small \textbf{\ours}                 & C \& P                  & D \& L                 & \textbf{\num{81.57894737}} \\
        \small \phantom{0}- \textit{direct code exec.}             & P                          & L                           & {\num{69.07894737}} \\
        \hline
        \small \texttt{Llama-3.1-8B-Inst}    & -                                   & -                                   & {\num{34.01315789}} \\
        \small TestChain             & C                                & D                              & {\num{46.27192982}} \\
        \small \textbf{\ours}                 & C \& P                  & D \& L                 & \textbf{{\num{53.28947368}}} \\
        \small \phantom{0}- \textit{direct code exec.}             & P                          & L                           & {\num{34.38596491}} \\
        \hline
        \small \texttt{Llama-3.1-70B-Inst}   & -                                   & -                                   & {\num{64.47368421}} \\
        \small TestChain             & C                                & D                              & {\num{71.49122807}} \\
        \small \textbf{\ours}                 & C \& P                  & D \& L                 & \textbf{{\num{78.28947368}}} \\
        \small \phantom{0}- \textit{direct code exec.}             & P                          & L                           & {\num{66.66666667}} \\

        \hline
        \small \texttt{Mistral-Large}            & -                                   & -                                   &  {\num{59.21052632}} \\
        \small TestChain             & C                                & D                              & {\num{55.70175439}} \\
        \small \textbf{\ours}                 & C \& P                  & D \& L                 & \textbf{{\num{67.10526316}}} \\
        \small \phantom{0}- \textit{direct code exec.}             & P                          & L                           & {\num{46.31578947}} \\

        \thickhline
    \end{tabular}}
    \end{flushleft}
    \caption{
        Pass@$1$ comparison on the LiveCodeBench test output prediction benchmark (Jan 1–Apr 1, 2024). Functional majority voting is applied to all baselines by default. {Each method is labeled by its grounding type, code (C) or pseudocode (P), and execution type, direct (D) or LLM-based (L).}
    }
    \label{tab:cutoff}
\end{table}

%% file: table/code_filtering.tex
\begin{table*}[htb!]
    \centering
    \small
    \renewcommand{\arraystretch}{1.2}
    \begin{tabular}{lll >{\color{black}}c >{\color{black}}c  >{\color{black}}c >{\color{black}}c}
        \thickhline
        \multirow{1}{*}{\textbf{Method}}  & \multirow{1}{*}{\textbf{Grounding}} & \multirow{1}{*}{\textbf{Execution}} & \textbf{Pass@$1$} & \textbf{Pass@$2$} & \textbf{Pass@$5$} & \textbf{Pass@$10$} \\
        \hline
        \small No Filtering & -                                   & -                                   & {\num{33.9}}   & {50.9}  & {64.8} & {73.3}          \\
        \small \textsc{CodeT}  & -                                   & -                                   & {54.2}  & {{62.8}}  & {{71.3}} & {74.3}         \\
        \small TestChain   & Code                                & Direct                              & {48.6}    & {54.9}  & {61.4} & {67.6}           \\
        \small \textbf{\ours}       & Code \& Pseudocode                  & Direct \& LLM-based                 & {\textbf{57.4}}    & {\textbf{65.2}}  & {\textbf{71.8}} & {\textbf{75.1}}           \\
        \small \phantom{0}- \textit{direct code exec.}   & Pseudocode                          & LLM-based                           & {{55.9}}   & {{64.5}}  & {\textbf{71.8}} & {\textbf{75.1}}           \\
        \thickhline
    \end{tabular}
    \caption{
    End-to-end code generation performance of \texttt{Llama-3.1-8B-Instruct} under different test output prediction methods, evaluated on LiveCodeBench-Easy (Jan 1, 2024–Feb 1, 2025). 
    }
    \label{tab:code_filtering}
\end{table*}

%% file: table/design_space.tex
\begin{table}[tb!]
    \centering
    \small
    \renewcommand{\arraystretch}{1.2}
    \resizebox{1.0\linewidth}{!}{
    \begin{tabular}{rcc}
        \thickhline
        \textbf{Grounding} & \textbf{Direct Exec.} & \textbf{LLM-based Exec.} \\
        \hline
        \textbf{Code} &  \cellcolor{my_pink!15} \textcolor{my_brown}{TestChain} & \textcolor{my_blue}{LLM-based Code Exec.} \\
        \textbf{Pseudocode} & - &  \cellcolor{my_pink!15}\textcolor{my_green}{LLM-based Pseudo. Exec.} \\
        \thickhline
    \end{tabular}}
    \caption{
        Design space taxonomy, categorized by
        execution target (code vs. pseudocode) and an execution mode (direct vs. LLM-based). \textcolor{my_pink}{\ours}, marked as a pink diagonal box, chooses the best combination of \textcolor{my_brown}{direct code execution} and \textcolor{my_green}{LLM-based pseudocode execution}.
    }
    \label{tab:design_space}
\end{table}

%% file: table/combination.tex
\begin{table}[t]
    \centering
    \small
    \renewcommand{\arraystretch}{1.2}

    \begin{flushleft}
    \small
    \renewcommand{\arraystretch}{1.1}
    \resizebox{1.0\linewidth}{!}{
    \begin{tabular}{lllcccc}
        \thickhline
        \textbf{Method}          & \textbf{Ground.} & \textbf{Exec.} & \textbf{Pass@$1$}                                                                                                               \\
        \hline
        \small \textit{no grounding}   & -                                   & -                                   & {\num{74.2458521870286}}\\
        \small \textcolor{my_brown}{\textit{direct code exec.} (TestChain)}            & C                                & D                              & \textcolor{my_brown}{\num{75.4524886877828}} \\

        \textcolor{my_blue}{\textit{LLM-based code exec.}} & C & L & \textcolor{my_blue}{\num{72.24736048265461}} \\
        \small \textcolor{my_green}{\textit{LLM-based pseudocode exec.}}             & P                          & L                           & \textcolor{my_green}{\num{74.05731523378581}} \\        
        \small \textcolor{my_blue}{\textit{direct \& LLM-based code exec. }}                 & C                  & D \& L                 & \textcolor{my_blue}{\num{77.37556561085974}} \\        
        \small \textcolor{my_blue}{\textit{LLM-based code \& pseudo. exec. }}                 & C \& P                  & L                 & \textcolor{my_blue}{\num{73.64253393665159}} \\        
        \small \textcolor{my_pink}{\textbf{\textit{dual exec.} (\ours~- \textit{path-weighted FMV})}}                 & C \& P                  & D \& L                 & \textcolor{my_pink}{\textbf{\num{80.76923076923077}}} \\

        \thickhline
    \end{tabular}}
    \end{flushleft}
    \caption{
        Test output prediction results of \texttt{GPT-4-Turbo-2024-04-09} across grounding combinations and execution methods on LiveCodeBench  (May 1, 2023–Apr 1, 2024). {Functional majority voting is applied across methods}. 
        We highlight representative baselines and our methods using distinct colors: 
        \textcolor{my_brown}{TestChain}, 
        \textcolor{my_green}{LLM-based pseudocode execution}, 
        and \textcolor{my_pink}{\ours}. 
        Variants that use \textcolor{my_blue}{LLM-based code execution} (alone or in combination) are marked in blue. {Each method is labeled by its grounding type, code (C) or pseudocode (P), and execution type, direct (D) or LLM-based (L).}
    }
    \label{tab:combination}
\end{table}

%% file: table/bigcodebench_hard_qwq.tex
\begin{table}[t]
    \begin{flushleft}
    \small
    \centering
    \arrayrulecolor{black}
    \begin{tabular}{>{\color{black}}l>{\color{black}}l>{\color{black}}l >{\color{black}}c}
        \toprule
        \textbf{Method}          & \textbf{Ground.} & \textbf{Exec.} & \color{black}{\textbf{Pass@$1$}} \\
        \midrule
        \small No Filtering       & -       & -       & {\num{24.04}} \\
        \small \texttt{CodeT}     & -       & -       & {\num{22.61}} \\
        \small TestChain          & C       & D       & {\num{24.04}} \\
        \small \textbf{\ours}      & C \& P  & D \& L  & \textbf{\num{25.34}} \\
        \small \phantom{0}- \textit{direct code exec.} & P & L & {\num{24.04}} \\
        \bottomrule
    \end{tabular}
    \end{flushleft}
    \caption{
        {End-to-end code generation performance of \texttt{QwQ-32B} under different test output prediction methods, evaluated on BigCodeBench-Hard.}
    }
    \arrayrulecolor{black}
    \label{tab:qwq32b_bigcodebench}
\end{table}

%% file: content/6_conclusion.tex
\section{Conclusion}
\label{sec:conclusion}

We proposed \ours, a dual-execution strategy that combines direct code execution and LLM-based pseudocode execution to enhance test output prediction.
By integrating two execution paths into a single decision framework, \ours mitigates both implementation errors and execution hallucinations.
Experiments on LiveCodeBench, BigCodeBench-Hard, and DevEval show that \ours sets a new state-of-the-art in test output prediction and enhances code generation via candidate filtering. 

\section*{Acknowledgment}
This work was supported by the Institute of Information \& communications Technology Planning \& Evaluation (IITP) grant funded by the Korea government(MSIT) (No.2022-0-00995, Automated reliable source code generation from natural language descriptions).

\section*{Limitations}
\label{sec:limitations}
Our analysis indicates that the effectiveness of direct vs. LLM-based execution varies with trace length, suggesting potential for adaptive strategies that predict the more reliable method per instance and adjust path weights in functional majority voting accordingly.
Our test input generation is intentionally simple; incorporating more diverse or diagnostic inputs such as edge cases may further benefit downstream applications like code generation.
The evaluation of newer models was limited by the current LiveCodeBench setup, which has not yet been updated to support recent versions for the test output prediction task.


%% file: content/Appendix.tex
\clearpage
\newpage

\input{content/5_related}

\input{table/prompt_engineering}

\section{{Analysis on Grounding Design and Reasoning Steps}}
\label{appendix:prompt_engineering}

{Beyond pseudocode design, we also considered alternative intermediate representations such as abstract syntax trees (ASTs), control flow graphs (CFGs), and type-based analyses. 
ASTs encode fine-grained syntactic details that obscure semantic intent, while type-based analyses offer limited structural context.
CFGs capture execution flow but impose a nontrivial burden on LLMs to linearize and interpret graph structures.
In contrast, pseudocode expressed in natural language offers a more accessible abstraction for reasoning-oriented models, while retaining sufficient alignment with program semantics.}

{As shown in Table~\ref{tab:prompt_engineering}, this design choice is empirically supported: 
moderately abstract pseudocode (Mid- or High-level) consistently outperforms overly concrete (Low-level) forms, 
suggesting that abstraction helps LLMs focus on semantic reasoning rather than token-level execution details.
Regarding reasoning steps, step-by-step and 3-step executions exhibit no consistent superiority across tasks, implying that the structure of reasoning matters less than the grounding it operates on.
Our default configuration (Mid-level pseudocode with step-by-step reasoning) achieves strong overall performance, while the best-performing setting (High-level + 3-step) provides only a marginal gain of 0.6 pp.
These observations suggest that grounding design plays a key role in stabilizing execution-based reasoning, while exploring richer graph- or type-aware forms remains an open direction.}

\section{On the Simplicity of the Method}
\label{appendix:method_simplicity}
Prior execution-based approaches such as TestChain rely solely on direct code execution, making them sensitive to code generation quality and prone to failure in end-to-end scenarios (\S\ref{appendix:why_testchain_bad}). \ours addresses this by adding an LLM-based pseudocode path, requiring no architectural changes or training overhead. Its simplicity is a practical advantage: the framework is easy to implement and extend with additional paths or weighting schemes.

\section{{Computational Overhead Analysis}}
\label{appendix:overhead}
{Under the functional majority voting with $n$ test outputs per input, the three execution paths differ mainly in the number of LLM calls and code executions.}
{First, direct code execution initially generates $n$ pseudocode candidates and their corresponding programs then executes each once, requiring $2n$ LLM calls and $n$ code executions.}
{Second, LLM-based pseudocode execution generates $n$ pseudocode candidates and evaluates them via the LLM, leading to $2n$ LLM calls without external execution.}
{Last, \ours performs both executions for the same $n$ candidates,\footnote{\edit{See Appendix~\ref{appendix:common_mode_failure} for a discussion on common mode failure mitigation.}} totaling $4n$ LLM calls and $n$ code executions.}

{Despite higher nominal cost, all methods can batch their operations: direct and LLM-based paths handle $n$ items per forward pass, while \ours processes $2n$.
Hence, with full parallelization, the effective runtime reduces to two forward passes for all methods.}

\section{Implementation Details}
\label{subsec:implementation}

\input{table/main}
\input{table/cutoff_orig_split}

\paragraph{Temperature}
For LLM inference, we use two different decoding settings:
When using greedy decoding, we set \texttt{temperature=0.0} and \texttt{top\_p=1.0}.
For nucleus sampling, we set \texttt{temperature=0.8} and \texttt{top\_p=0.95}.
Specifically, pseudocode generation is performed using the nucleus sampling, while code generation and LLM-based Execution uses greedy decoding.  For \textbf{No Grounding}, outputs are directly predicted by nucleus sampling.

\paragraph{Path-Weighted FMV}
To reflect path-level confidence, we assign a higher weight $w_{{high}}{=}2$ to predictions when all the outputs of a given path agree unanimously; otherwise, we use the base weight $w_{{base}}{=}1$.

\paragraph{Fallback} As aforementioned, LLM-based execution uses nucleus sampling with high temperature to encourage diversity.
To avoid committing to uncertain predictions, we fall back to a default ungrounded prediction when the LLM-based execution outputs from this path are not unanimous. 

\paragraph{Method-Specific Hyperparameters}
\begin{itemize}
    \item \textbf{No Grounding}: Directly predicts $n{=}10$ outputs from the problem and input.
    \item \textbf{TestChain}~\cite{li2024largelanguagemodelstest}: Generates $l{=}10$ outputs via direct code execution in Section~\ref{sec:proposed_main}.
    \item \textbf{LLM-based Pseudocode Execution}: Generates $m{=}10$ outputs via LLM-based pseudocode execution in Section~\ref{sec:proposed_main}.
    \item \textbf{\ours}: Combines both strategies, generating $l{=}10$ outputs via direct execution and $m{=}10$ via LLM-based execution. 
\end{itemize}
\edit{For end-to-end code generation, we used $l{=}m{=}5$ in \ours while keeping the baseline settings unchanged.}

\paragraph{\edit{Implementation of TestChain}}
\edit{Given the absence of a public implementation for TestChain, we reproduced the method with a specific design choice to ensure a competitive baseline. The original TestChain relies on a ReAct~\cite{react}-style interaction loop for incremental code construction, where new code segments are appended without modifying previously generated parts. However, prior studies~\cite{jiang2024self} indicate that such multi-turn incremental strategies often underperform compared to single-pass generation. Consequently, we adopted a direct-execution formulation that synthesizes the complete solution in a single pass, avoiding the potential limitations of the incremental approach and providing a more robust standard for comparison.}

\paragraph{LLM Inference Backends}
We utilize multiple inference backends depending on availability and deployment settings:
(1) {vLLM}~\cite{kwon2023efficient}: An optimized inference engine supporting fast batched decoding;
(2) {Ollama}~\cite{ollama2025docs}: A lightweight on-device LLM inference runtime;
(3) {Snowflake Cortex}~\cite{snowflake2024cortex}: A commercial LLM inference platform;
(4) {OpenAI API}~\cite{openai2023api}: Used for models such as \texttt{GPT-4-Turbo-2024-04-09}.
All models are used without quantization, employing full-precision FP16 weights when possible.

We run local inference with vLLM or Ollama on eight NVIDIA RTX 3090 GPUs. Under this setup, \texttt{Llama-3.1-8B-Instruct} processes the full LiveCodeBench benchmark ($\sim$ 22K tokens/min) in about one hour, while \texttt{Llama-3.1-70B-Instruct} achieves ~4K tokens/min, taking roughly six hours.

\input{table/code_filtering_humanevalplus}
\paragraph{Prompt Orchestration} We implement the system pipeline using the {LangChain} framework~\cite{langchain2022}, which provides modular abstractions for prompt orchestration and LLM interaction.
For managing multi-step workflows and maintaining intermediate states across executions, we adopt {LangGraph}~\cite{langgraph2023},
a directed-graph-based orchestration engine built on top of LangChain.
This allows flexible integration of both code-based and pseudocode-based reasoning within a unified execution graph.

\begin{figure*}[t]
    \centering
    \includegraphics[width=0.8\linewidth]{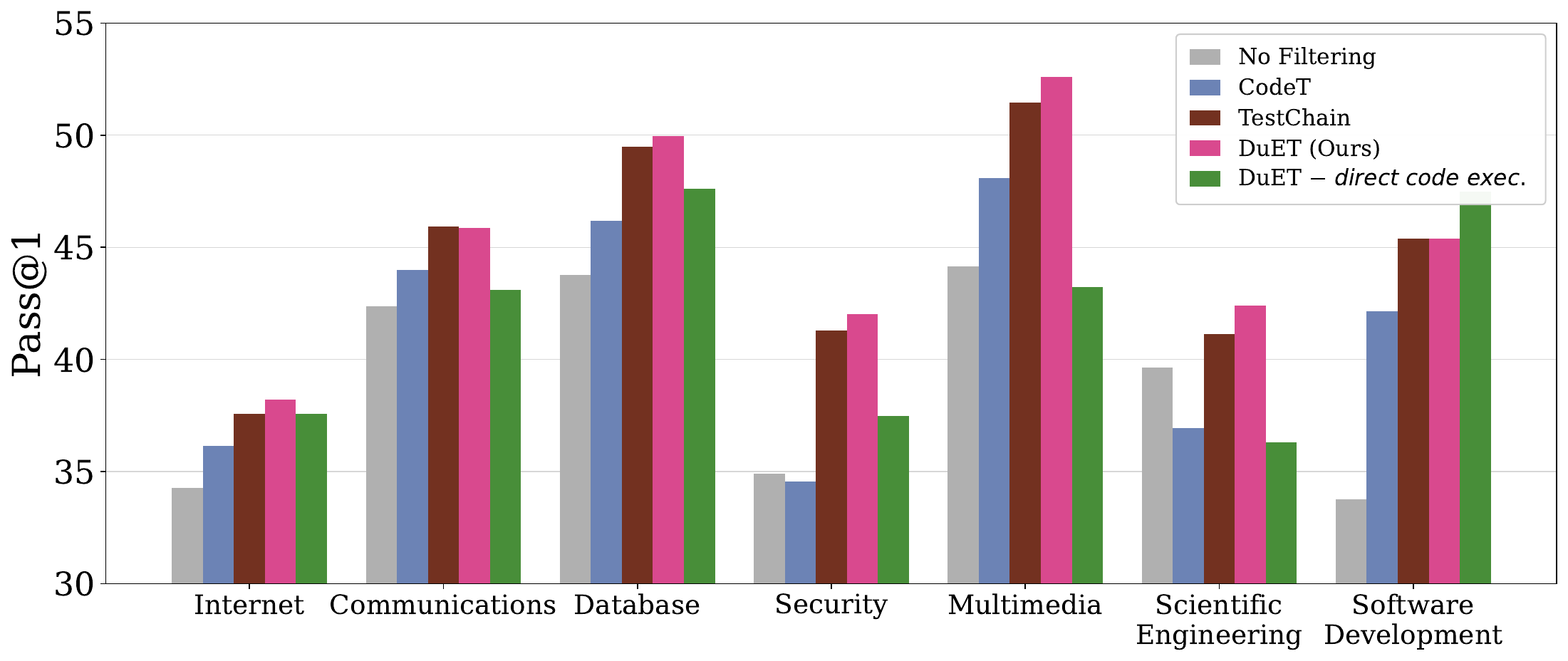}
    \caption{
        \edit{End-to-end code generation performance of \texttt{Llama-3.1-8B-Instruct} evaluated on DevEval.}
    }
    \label{fig:repo-level_full}
\end{figure*}

\section{\edit{\textbf{Repository-level Code Generation}}}
\label{appendix:repo-level}
\edit{For the DevEval~\cite{li-etal-2024-deveval} evaluation, we utilized Llama-3.1-8B-Instruct to generate 20 code candidates and test cases per problem.
To strictly assess the discriminative power of code filtering, we excluded instances where all candidates were either completely correct or incorrect, as these scenarios do not differentiate filtering performance.
Consequently, the evaluation was conducted on a subset of 367 problems selected from the original 1,825 problems across 10 domains.
Other configurations follow the settings described in Section~\ref{sec:code_generation}.}

\section{{Difficulty Categorization in Test Output Prediction}}
\label{appendix:input_difficulty}
{In terms of difficulty categorization, test output prediction is better characterized by the test input, since even problems with difficult reference solutions may present inputs that are easy to predict. 
Thus, we redefine difficulty for this task based on the test input rather than the problem: specifically, we generate 10 code samples per test input using \texttt{GPT-4-Turbo-2024-04-09} (the strongest baseline) and compute the proportion of samples that produce the correct output. }

{As shown in Table~\ref{tab:main}, direct code execution (TestChain) outperforms LLM-based pseudocode execution (\ours - \textit{direct code exec.}) on the Easy and Medium subsets, likely due to its precision on straightforward inputs. Conversely, LLM-based pseudocode execution excels on the Hard subset, where more challenging inputs benefit from additional reasoning. The dual execution approach (\ours) achieves the best overall performance: it ranks highest on the Easy and Medium subsets, and second-best on the Hard subset.
}

\edit{These results confirm that the two paths cover each other's weaknesses. For Easy and Medium inputs, the direct-execution path remains effective because even imperfectly generated code often yields correct outputs for simpler cases. In contrast, Hard inputs require strict logical correctness, making the pseudocode-based simulation path more critical. \ours ranks 1st on the All, Easy, and Medium splits, and 2nd on Hard, while consistently outperforming the direct-execution baseline.}

{We also report the results under the original problem-based difficulty split in Table~\ref{tab:cutoff_orig_split} for comparison. Interestingly, under this split, TestChain achieves the highest performance on the Hard subset, contrary to its underperformance on the Hard subset defined by input difficulty (Table~\ref{tab:main}). This discrepancy highlights the limitations of using problem-level metadata, such as reference code complexity, as a proxy for prediction difficulty. In contrast, input-based categorization more directly reflects the reasoning challenges encountered during test output prediction, better aligning with the actual performance characteristics of execution-based methods.}
\input{table/cutoff_exaone}

{Meanwhile, Table~\ref{tab:qwq32b_bigcodebench} shows that with a stronger reasoning model QwQ-32B~\cite{qwq32b}, both code and pseudocode executions become more reliable, enabling \ours to realize its advantage even on inputs categorized as \textsc{Hard}. 
This suggests that input-based difficulty interacts with model capability: as reasoning strength increases, brittle failure modes in direct execution diminish, narrowing method gaps and amplifying the gains of dual execution.}

\input{table/consensus_exaone}
\input{table/model_cutoff_info}

\section{{End2End Code Generation on HumanEval and HumanEval+}}
\label{appendix:end2end_humanevalplus}
{Table~\ref{tab:code_filtering_humanevalplus} reports end-to-end code generation results on HumanEval~\cite{codex} and HumanEval+~\cite{YourCodeGenerated}.
\ours achieves the highest Pass@$1$ on both benchmarks, followed closely by LLM-based pseudocode execution (\ours~–~\textit{direct code exec.}).
These are the only two methods that outperform the base code filtering approach (\textsc{CodeT}), while TestChain again underperforms relative to \textsc{CodeT}, consistent with observations on Table~\ref{tab:code_filtering}.
These results confirm that \ours improves performance not only on challenging tasks like LiveCodeBench but also on easier benchmarks such as HumanEval(+), where overall model accuracy is already high.
The relatively small performance gains are likely due to the saturated nature of these benchmarks, where most candidate solutions are already correct, making the benefits of better test output prediction marginal.}

\section{\edit{Test Output Prediction Results with Contamination Risk Models}}
\label{appendix:exaone}
\edit{Table~\ref{tab:cutoff_exaone} presents the performance of models with potential data contamination, excluded from the main text.
The knowledge cutoff dates for all models, including these, are detailed in Table~\ref{tab:model_cutoff_dates}.
Overall, the results show performance trends consistent with Table~\ref{tab:cutoff} in the main context.}


\section{{Path-Weighted FMV: Safe Gains at No Cost}}
\label{appendix:consensus}
{Table~\ref{tab:consensus_exaone} shows that path-weighted FMV consistently matches or slightly outperforms the standard FMV across all models on LiveCodeBench. 
While the gains are modest (up to 0.7 pp Pass@$1$), the method requires \edit{zero additional cost} and ensures \edit{non-decreasing accuracy}. 
{We leave more advanced weighting mechanisms for future work (see Appendix~\ref{appendix:future_work}).}

\section{\edit{Discussion}}
\subsection{\edit{Difference between Test Output Prediction and Code Execution}}
\label{appendix:cruxeval}
\edit{While both tasks involve predicting the result of a program, there is a fundamental distinction regarding the information provided to the model. Code execution tasks, exemplified by benchmarks such as CRUXEval~\cite{pmlr-v235-gu24c}, provide the ground-truth code and evaluate a model's ability to simulate the execution process (i.e., predicting outputs given specific code and inputs). In contrast, the test output prediction task in LiveCodeBench operates without access to ground-truth code. It requires the model to derive the intended program behavior solely from a natural language problem description. Consequently, this task evaluates the model's capability to reason about requirements and synthesize logic from scratch, mimicking the real-world software development process, rather than the narrower capability of mentally tracing provided code.}

\subsection{\edit{Common Mode Failure Mitigation}}
\label{appendix:common_mode_failure}
\edit{In our dual-path framework, common mode failure refers to the risk where a single upstream error in the LLM-generated pseudocode propagates to both execution paths, causing simultaneous failure. \ours mitigates this by diversifying the pseudocode space, generating multiple candidates rather than relying on a single interpretation. Because the two paths have different error profiles, a flawed pseudocode is unlikely to cause both paths to fail in the same way, preventing upstream errors from dictating the final prediction.}

\subsection{Metric Comparison Justification}
\label{appendix:metric_comparison}
\edit{In Table~\ref{tab:leaderboard}, we compare our ranking-based results against the official leaderboard's unbiased Pass@$k$. Since unbiased Pass@$k$ represents the expected performance of a {random selection baseline} (i.e., uniform random sampling), this comparison quantifies the effectiveness of our \textit{intelligent selection} over random selection, consistent with prior work on neural code rankers~\cite{inala2022faultaware,li2025toward, lahiri2022interactive}.}

\section{{Future Directions}}
\label{appendix:future_work}
{
A promising future extension of \ours is to develop an adaptive weighting mechanism that dynamically adjusts the contribution of each execution path based on its estimated reliability or difficulty.
In particular, a lightweight model could be trained to predict which path (code or pseudocode) is more trustworthy for a given instance, using signals such as the log-likelihood of generated code, execution-trace length, or recent error patterns.
Such a learned weighting scheme would generalize our current unanimous rule, allowing \ours to assign path weights more selectively based on task difficulty and reasoning complexity.
}

\section{Case Study (generated by \texttt{GPT-4-Turbo-2024-04-09})}

\subsection{Example of Execution Hallucination (Figures~\ref{fig:problem_description}-\ref{case:execution_hallucination_code})}
\label{appendix:llm_trace_error}

In Figure~\ref{case:execution_hallucination}, lines \texttt{4--6} of the LLM-based execution output reflect a hallucinated assumption about the input string \texttt{``2245047"}: it incorrectly identifies two zeros at indices 5 and 6, while in fact, only a single \texttt{`0'} exists at index 5. As a result, forming a suffix ending in \texttt{``00"} is impossible. This leads to an invalid computation of the operation cost and ultimately an incorrect prediction that no deletions are required. By contrast, when directly executing the correctly generated code shown in Figure~\ref{case:execution_hallucination_code}, the reasoning proceeds without such hallucinations and yields the correct output.


\begin{figure*}[t]
\centering
\begin{WideMinted}[fontsize=\tiny, frame=lines]
You are given a 0-indexed string num representing a non-negative integer.
In one operation, you can pick any digit of num and delete it. Note that if you delete all the digits of num, num becomes 0.
Return the minimum number of operations required to make num special.
An integer x is considered special if it is divisible by 25.
\end{WideMinted}
\caption{Problem description.}
\label{fig:problem_description}
\end{figure*}

\begin{figure*}[t]
\centering
\begin{WideMinted}[fontsize=\tiny, frame=lines]
Algorithm minimumOperations(num):
    1. Initialize a list possible_ends with values ["00", "25", "50", "75"] which represent the numbers divisible by 25.
    2. Set min_operations to a large value (e.g., length of num + 1) to keep track of the minimum operations needed.
    3. Iterate over each possible_end in possible_ends:
        a. Initialize two pointers, i and j, to the end of the string num.
        b. Set found_digits to 0, representing how many digits of possible_end have been matched.
        c. While iterating from the end of num towards the beginning:
            - If found_digits is 0 and num[i] matches the second character of possible_end:
                * Increment found_digits to 1.
                * Move pointer j to i.
            - Else if found_digits is 1 and num[i] matches the first character of possible_end:
                * Increment found_digits to 2.
                * Break the loop.
            - Decrement i.
        d. If found_digits is 2:
            - Calculate the operations needed as (j - i - 1) + (length of num - j - 1).
            - Update min_operations with the minimum of itself and the calculated operations.
    4. If min_operations is still set to its initial large value, return the length of num - 1 (all digits removed except one).
    5. Return min_operations.
\end{WideMinted}
\caption{Generated Pseudocode \textcolor{my_correct}{(Correct)}.}
\end{figure*}

\begin{figure*}[t]
\centering
\begin{WideMinted}[fontsize=\tiny, frame=lines, highlightlines={4,5,6}, highlightcolor=red!20]
To determine the minimum number of operations required to make the number "2245047" special (divisible by 25), we need to find the fewest deletions needed to end the number with "00", "25", "50", or "75".

1. For "00":
   - The last '0' is at index 6.
   - The second last '0' is at index 5.
   - Operations needed: (6 - 5 - 1) + (7 - 6 - 1) = 0 + 0 = 0.

2. For "25":
   - The last '5' is at index 4.
   - The nearest '2' before that '5' is at index 2.
   - Operations needed: (4 - 2 - 1) + (7 - 4 - 1) = 1 + 2 = 3.

3. For "50":
   - The last '0' is at index 6.
   - The nearest '5' before that '0' is at index 4.
   - Operations needed: (6 - 4 - 1) + (7 - 6 - 1) = 1 + 0 = 1.

4. For "75":
   - The last '5' is at index 4.
   - There is no '7' before that '5'.

The minimum operations among these options are 0 (for ending "00"). Thus, no deletions are needed to make the number special.

```
0
```
\end{WideMinted}
\caption{LLM-based Execution \textcolor{red!50}{(Execution Hallucination)}.}
\label{case:execution_hallucination}
\end{figure*}

\begin{figure*}[t]
\centering
\begin{WideMinted}[fontsize=\tiny, frame=lines]
def minimumOperations(num: str) -> int:
    possible_ends = ["00", "25", "50", "75"]
    min_operations = len(num) + 1
    
    for possible_end in possible_ends:
        i = len(num) - 1
        found_digits = 0
        j = 0
        
        while i >= 0:
            if found_digits == 0 and num[i] == possible_end[1]:
                found_digits = 1
                j = i
            elif found_digits == 1 and num[i] == possible_end[0]:
                found_digits = 2
                break
            i -= 1
        
        if found_digits == 2:
            operations = (j - i - 1) + (len(num) - j - 1)
            min_operations = min(min_operations, operations)
    
    if min_operations == len(num) + 1:
        return len(num) - 1  # All digits removed except one
    
    return min_operations
\end{WideMinted}
\caption{Generated Code \textcolor{my_correct}{(Correct)}.}
\label{case:execution_hallucination_code}
\end{figure*}

\subsection{Example of Implementation Error (Figures~\ref{fig:fig13_start}-\ref{case:implementation_error})}
\label{appendix:implementation_error}
Figure~\ref{case:implementation_error} illustrates a translation discrepancy. Line 6 of the pseudocode (Figure~\ref{case:implementation_error_pseudo}) implicitly requires sorting the indices of each connected component before reinsertion, but the implementation in lines 39–41 fails to do so.
In the code, the list of indices is taken directly from the BFS discovery order (\texttt{indices = component}) and paired with sorted values using zip. However, since the BFS order is arbitrary, smaller values may be placed at larger indices, and the resulting array is not guaranteed to be lexicographically minimal. To align with the pseudocode's intent, the index list must be explicitly sorted (e.g., replacing line 39 with \texttt{indices = sorted(component)}) to ensure the smallest values occupy the leftmost positions and preserve lexicographic order. In contrast, as shown in Figure~\ref{case:implementation_error_llm_execution}, LLM-based pseudocode execution correctly infers this ordering requirement and predicts the correct output while bypassing low-level implementation details.


\begin{figure*}[t]
\centering
\begin{WideMinted}[fontsize=\tiny, frame=lines]
You are given a 0-indexed array of positive integers nums and a positive integer limit.
In one operation, you can choose any two indices i and j and swap nums[i] and nums[j] if |nums[i] - nums[j]| <= limit.
Return the lexicographically smallest array that can be obtained by performing the operation any number of times.
An array a is lexicographically smaller than an array b if in the first position where a and b differ, array a has an element that is less than the corresponding element in b. For example, the array [2,10,3] is lexicographically smaller than the array [10,2,3] because they differ at index 0 and 2 < 10.
\end{WideMinted}
\caption{Problem Description.}
\label{fig:fig13_start}
\end{figure*}

\begin{figure*}[t]
\centering
\begin{WideMinted}[fontsize=\tiny, frame=lines, highlightlines={}, highlightcolor=black!20]
Algorithm lexicographicallySmallestArray(nums, limit):
    1. Initialize a graph with nodes representing each index of the array nums.
    2. Connect two nodes with an edge if the difference between their corresponding values in nums is less than or equal to limit.
    3. Perform a Depth-First Search (DFS) or Breadth-First Search (BFS) to find all connected components of the graph.
    4. For each connected component:
       - Extract the subarray of nums corresponding to the indices in this component.
       - Sort this subarray to get the lexicographically smallest order.
    5. Replace the elements in the original nums array at the positions of the connected component with the sorted values.
    6. Return the modified nums array as the result.
\end{WideMinted}
\caption{Generated Pseudocode \textcolor{my_correct}{(Correct)}.}
\label{case:implementation_error_pseudo}
\end{figure*}

\begin{figure*}[t]
\centering
\begin{WideMinted}[fontsize=\tiny, frame=lines]
To solve the problem, we need to find the lexicographically smallest array by swapping elements within a certain limit of their differences. The approach involves creating a graph where each index in the array represents a node, and an edge exists between nodes if the absolute difference between their corresponding values is less than or equal to the given limit.

For the given input:
- nums = [1, 5, 3, 9, 8]
- limit = 2

We analyze the differences:
- |1 - 5| = 4 (greater than 2, no edge)
- |1 - 3| = 2 (equal to 2, edge between indices 0 and 2)
- |1 - 9| = 8 (greater than 2, no edge)
- |1 - 8| = 7 (greater than 2, no edge)
- |5 - 3| = 2 (equal to 2, edge between indices 1 and 2)
- |5 - 9| = 4 (greater than 2, no edge)
- |5 - 8| = 3 (greater than 2, no edge)
- |3 - 9| = 6 (greater than 2, no edge)
- |3 - 8| = 5 (greater than 2, no edge)
- |9 - 8| = 1 (less than 2, edge between indices 3 and 4)

From this, we can identify connected components:
- Component 1: Indices [0, 2, 1] (values [1, 3, 5])
- Component 2: Indices [3, 4] (values [9, 8])

Sorting each component:
- Component 1 sorted: [1, 3, 5]
- Component 2 sorted: [8, 9]

Reconstructing the array with sorted components:
- [1, 3, 5, 8, 9]

Thus, the lexicographically smallest array that can be obtained is [1, 3, 5, 8, 9].
```
[1, 3, 5, 8, 9]
```
\end{WideMinted}
\caption{LLM Execution \textcolor{my_correct}{(Correct)}.}
\label{case:implementation_error_llm_execution}
\end{figure*}

\begin{figure*}[t]
\centering
\begin{WideMinted}[fontsize=\tiny, frame=lines, highlightlines={39}, highlightcolor=black!20]
def lexicographicallySmallestArray(nums: List[int], limit: int) -> List[int]:
    from collections import defaultdict, deque
    
    n = len(nums)
    graph = defaultdict(list)
    
    # Build the graph
    for i in range(n):
        for j in range(i + 1, n):
            if abs(nums[i] - nums[j]) <= limit:
                graph[i].append(j)
                graph[j].append(i)
    
    visited = [False] * n
    components = []
    
    # Function to perform BFS and find connected components
    def bfs(start):
        queue = deque([start])
        component = []
        visited[start] = True
        while queue:
            node = queue.popleft()
            component.append(node)
            for neighbor in graph[node]:
                if not visited[neighbor]:
                    visited[neighbor] = True
                    queue.append(neighbor)
        return component
    
    # Find all connected components
    for i in range(n):
        if not visited[i]:
            component = bfs(i)
            components.append(component)
    
    # Sort each component and place back into nums
    for component in components:
        indices = component
        values = [nums[i] for i in indices]
        values.sort()
        for idx, val in zip(indices, values):
            nums[idx] = val
    
    return nums
\end{WideMinted}
\caption{Generated Code \textcolor{black!50}{(Implementation Error)}.}
\label{case:implementation_error}
\end{figure*}

\section{Prompt Templates}
We adopt task-specific prompt templates for each stage of our pipeline. In Figures~\ref{lst:pseudocode-prompt}-\ref{lst:testcase-gen-prompt}, we list the prompt formats used for each.
All templates are written in Jinja2 format.

\begin{figure*}[h]
\begin{WideMinted}[fontsize=\tiny, frame=lines]
# Inputs
## Problem
```plaintext
{{ problem }}
```

## Starter code
```python
{{ starter_code }}
```

# Instruction
I provided you a coding problem.
You need to write a pseudocode for the problem.
Here are some conditions.
- The output should be written in a separate code block using three backticks (```) at the beginning and end.
- The part surrounded by double curly braces in the output format below is a placeholder. You need to replace it with an appropriate value to generate the output.
- Print only one code block. Do not print any other code block.

# Output Format{
```plaintext
{{ pseudocode }}
```{
\end{WideMinted}
\caption{Prompt template for pseudocode generation.}
\label{lst:pseudocode-prompt}
\end{figure*}

\begin{figure*}[h]    
\begin{WideMinted}[fontsize=\tiny, frame=lines]
# Inputs
## Problem
```plaintext
{{ problem }}
```

## Starter code
```python
{{ starter_code }}
```

## Pseudocode
```plaintext
{{ pseudocode }}
```

# Instruction
You need to write a python solution for the problem.
Here are some conditions.
- The code should implement the same algorithm as the given pseudocode.
- The output should be written in a separate code block using three backticks (```) at the beginning and end.
- Do not include `self` in the function input arguments.
- The part surrounded by double curly braces in the output format is a placeholder. You need to replace it with an appropriate value to generate the output.
- Print only one code block. Do not print any other code block.

# Output Format{
```python
{{ code }}
```{
\end{WideMinted}
\caption{Prompt template for code generation.}
\label{lst:code-prompt}
\end{figure*}
\newpage
\begin{figure*}[h]
\begin{WideMinted}[fontsize=\tiny, frame=lines]
# Inputs
## Problem
```plaintext
{{ problem }}
```

## Testcase Input
```
{{ tc_input }}
```

# Instruction
You need to predict the output of the testcase input based on the provided coding problem.
Here are some conditions.
- Write the reasoning steps before providing the final answer.
- Do not write the correct code or code to execute the test case; instead, write the expected output.
- The final output should not be in natural language format but should consist of the output itself only.
- The part surrounded by double curly braces in the output format below is a placeholder. You need to replace it with an appropriate value to generate the output.
- The final output should be written in a code block using three backticks (```) at the beginning and end.

# Output Format{
{{ reasoning }}
```json
{{ output }}
```{
\end{WideMinted}
\caption{Prompt template for LLM-based execution (no grounding).}
\label{lst:llm-exec-prompt-noground}
\end{figure*}

\begin{figure*}[h]
\begin{WideMinted}[fontsize=\tiny, frame=lines]
# Inputs
## Problem
```plaintext
{{ problem }}
```

## Starter code
```python
{{ starter_code }}
```

## Pseudocode
```plaintext
{{ pseudocode }}
```

## Testcase Input
```plaintext
{{ tc_input }}
```

# Instruction
Predict the testcase output that would result from executing the given testcase input on the provided problem.
Here are some conditions.
- Before arriving at the final answer, understand the algorithm described in the pseudocode and write reasoning that explains the process of deriving the final answer based on the given testcase input.
- The final output should not be in natural language format but should consist of the output itself only.
- The part surrounded by double curly braces in the output format below is a placeholder. You need to replace it with an appropriate value to generate the output.
- Your final expected output should always be placed at the end of your response, enclosed by a pair of ```.

# Output Format{
{{ reasoning }}
```
{{ expected_output }}
```{
\end{WideMinted}
\caption{Prompt template for LLM-based execution (pseudocode grounding).}
\label{lst:llm-exec-prompt-pseudocode}
\end{figure*}

\begin{figure*}[h]    
\begin{WideMinted}[fontsize=\tiny, frame=lines]
# Inputs
## Problem
```plaintext
{{ problem }}
```

## Starter code
```python
{{ starter_code }}
```

## Code
```python
{{ pseudocode }}
```

## Testcase Input
```plaintext
{{ tc_input }}
```

# Instruction
Predict the testcase output that would result from executing the given testcase input on the provided problem.
Here are some conditions.
- Before arriving at the final answer, understand the algorithm described in the code and write reasoning that explains the process of deriving the final answer based on the given testcase input.
- The final output should not be in natural language format but should consist of the output itself only.
- The part surrounded by double curly braces in the output format below is a placeholder. You need to replace it with an appropriate value to generate the output.
- Your final expected output should always be placed at the end of your response, enclosed by a pair of ```.

# Output Format{
{{ reasoning }}
```
{{ expected_output }}
```{
\end{WideMinted}
\caption{Prompt template for LLM-based execution (code grounding).}
\label{lst:llm-exec-prompt-code}
\end{figure*}
\clearpage
\begin{figure*}[h]    
\begin{WideMinted}[fontsize=\tiny, frame=lines]
# Inputs
## Problem
```plaintext
{{ problem }}
```

## Starter Code
```python
{{ starter_code }}
```

# Instruction
You need to generate testcase inputs for the problem.
Here are some conditions.
- Provide reasoning before generating the main output.
- Write exactly 3 test case inputs.
- If the test case input format supports executing multiple test cases at once, ensure each test case tests only a single scenario.
- Do not generate test case outputs.
- If starter code is provided, format the input as a JSON string representing a dictionary where the keys correspond to the input argument names of the given function.
- If starter code is not provided, format the input as expected by the program via standard input (stdin) so it can be directly used as input.
- Replace placeholders surrounded by double curly braces in the output format with appropriate values.
- Continue writing test case inputs up to a total of 3, following the provided format.
- Enclose all outputs in separate code blocks using three backticks (```).

# Output Format{
## Reasoning
{{ reasoning }}
## Test Case Inputs
### Test Case Input 1
```
{{ testcase input 1 }}
```
### Test Case Input 2
```
{{ testcase input 2 }}
```
### Test Case Input 3
```
{{ testcase input 3 }}
```
{
\end{WideMinted}
\caption{Prompt template for test input suite generation.}
\label{lst:testcase-gen-prompt}
\end{figure*}

\twocolumn

\onecolumn

%% file: content/5_related.tex
\section{{Related Work}}
\label{sec:related}

\paragraph{{Grounded Test Output Prediction}}
{Test case generation provides input-output pairs to verify code correctness~\cite{chen2023codet,huang-etal-2024-enhancing,han-etal-2024-archcode}, while also enabling feedback-driven refinement~\cite{shinn2023reflexion,he-etal-2024-cocost}.  
Recent methods~\cite{LiveCodeBench,li2024largelanguagemodelstest} emphasize {test output prediction} as a distinct subtask, where the goal is to infer the correct output given a problem and test input, often requiring nontrivial program reasoning.  
Whereas \textsc{CodeT}~\cite{chen2023codet} couples input and output generation, TestChain~\cite{li2024largelanguagemodelstest} explicitly decouples them and performs grounded prediction by executing code generated from the problem description, highlighting the need for reliable grounding to support output prediction.}



\paragraph{{Intermediate Representations for Code Reasoning}}
{A parallel line of work explores using \textbf{intermediate representations}, such as pseudocode or high-level plans, to bridge natural language and executable code.  
Plan-first methods like~\citet{huang2023codecot},~\citet{jiang2024self}, and~\citet{islam2024mapcoder} generate action sketches before code, improving alignment with the problem intent.  
SCoT~\cite{li2025structured} introduces structured intermediate forms, and CodeChain~\cite{le2023codechain} further modularizes the process into reusable components.  
By abstracting away low-level syntax, these works suggest that intermediate representations allow models to focus on semantic intent, facilitating more accurate and flexible reasoning. } 

\paragraph{{Our Distinction}}
{While \ours tackles test output prediction by grounding as TestChain does, it also overcomes the brittleness of direct execution by leveraging pseudocode as an intermediate representation.
By combining both paths, \ours pairs the symbolic precision of code (\S\ref{sec:analysis_reasoning}) with the abstraction of pseudocode (\S\ref{sec:analysis_code_correctness}), covering failure modes that neither path handles well alone. }

%% file: table/prompt_engineering.tex
\begin{table}[t]
    \centering
    \small
    \renewcommand{\arraystretch}{1.1}
    \begin{tabular}{>{\color{black}}l >{\color{black}}l >{\color{black}}c}
        \toprule
        \textbf{Pseudocode} & \textbf{LLM-based} & \multirow{2}{*}{\textbf{Pass@1}} \\
        \textbf{Granularity} & \textbf{Execution} &  \\
        \hline
        Low             & Step-by-step (default) & 31.1 \\
        Low             & 3-step                 & 31.3 \\\hdashline
        Mid (default)   & Step-by-step (default) & 34.4 \\
        Mid (default)   & 3-step                 & 32.5 \\\hdashline
        High            & Step-by-step (default) & 25.1 \\
        High            & 3-step                 & 35.0 \\
        \bottomrule
    \end{tabular}
    \caption{
        {Test output prediction performance comparison of \texttt{Llama-3.1-8B-Instruct} with LLM-based pseudocode execution over prompt design choices on LiveCodeBench (May 1, 2023–Apr 1, 2024).  Rows vary the granularity of the generated pseudocode (low, mid, high), while columns compare two LLM-based execution prompting strategies: a step-by-step simulation and a condensed 3-step reasoning format. Default settings are indicated in parentheses. Functional majority voting is applied with 10 independent samples per output.}
    }
    \arrayrulecolor{black}
    \label{tab:prompt_engineering}
\end{table}

%% file: table/main.tex
\begin{table*}[t]
    \centering
    \small
    \renewcommand{\arraystretch}{1.2}
    \begin{tabular}{>{\color{black}}l >{\color{black}}l >{\color{black}}l >{\color{black}}l >{\color{black}}c >{\color{black}}c >{\color{black}}c >{\color{black}}c}
        \toprule
        \multirow{2}{*}{\textbf{Model}}          & \multirow{2}{*}{\textbf{Grounding}} & \multirow{2}{*}{\textbf{Execution}} & \multicolumn{4}{c}{\multirow{1}{*}{{\textbf{Pass@$1$}}}}                                                                                                               \\

                                                 &                                     &                                     & \multirow{-1.1}{*}{\textbf{All}}                       & \multirow{-1.1}{*}{\textbf{Easy}} & \multirow{-1.1}{*}{\textbf{Medium}} & \multirow{-1.1}{*}{\textbf{Hard}} \\
        \hline
        \small \texttt{Llama-3.1-8B-Instruct}
                                                 & -                                   & -                                   & \num{34.01315789}                                & \num{35.28571429}           & \num{40.90909091}             & \num{23.68421053}          \\
        \small TestChain             & Code                                & Direct                              & {\num{46.27192982}}                                      & {\num{55.23809524}}         & {\num{51.51515152}}                   & {\num{23.68421053}}                 \\
        \small \ours                 & Code \& Pseudocode                  & Direct \& LLM-based                 & \textbf{\num{53.28947368}}                             & \textbf{\num{61.42857143}}         & \textbf{\num{62.5}}          & {\num{27.63157895}}        \\
        \small \phantom{0}- \textit{direct code exec.}             & Pseudocode                          & LLM-based                           & \num{34.38596491}                                      & \num{35.14285714}                 & \num{27.27272727}                   & \textbf{\num{41.22807018}}                 \\

        \bottomrule
    \end{tabular}
    \caption{
        {Test output prediction results on LiveCodeBench (Jan 1–Apr 1, 2024), where difficulty levels of test inputs are estimated using the correctness rate of code generated by the best performing model (\texttt{GPT-4-Turbo-2024-04-09}).
        Functional majority voting is applied to all baselines by default.    }}
    \arrayrulecolor{black}
    \label{tab:main}
\end{table*}

%% file: table/cutoff_orig_split.tex
\begin{table*}[htb!]
    \centering
    \small
    \renewcommand{\arraystretch}{1.2}
    \begin{tabular}{lll >{\color{black}}c >{\color{black}}c >{\color{black}}c >{\color{black}}c}
        \toprule
        \multirow{2}{*}{\textbf{Model}}          & \multirow{2}{*}{\textbf{Grounding}} & \multirow{2}{*}{\textbf{Execution}} & \multicolumn{4}{c}{\multirow{1}{*}{\textbf{Pass@$1$}}}                                                                                                               \\

                                                 &                                     &                                     & \multirow{-1.1}{*}{\color{black}\textbf{All}}                       & \multirow{-1.1}{*}{\color{black}\textbf{Easy}} & \multirow{-1.1}{*}{\color{black}\textbf{Medium}} & \multirow{-1.1}{*}{\color{black}\textbf{Hard}} \\
        \hline
        \small \texttt{Llama-3.1-8B-Instruct}    & -                                   & -                                   & \num{34.01315789}                                      & \num{39}                          & \num{35.25}                   & \num{18.18181818}                  \\
        \small TestChain             & Code                                & Direct                              & {\num{46.27192982}}                                      & \num{58}                          & {\num{39.58333333}}                   & \textbf{\num{43.93939394}}                 \\
        \small \ours                 & Code \& Pseudocode                  & Direct \& LLM-based                 & \textbf{\num{53.28947368}}                                      & \textbf{\num{64}}                 & \textbf{\num{51.25}}                   & {\num{36.36363636}}                 \\
        \small \phantom{0}- \textit{direct code exec.}             & Pseudocode                          & LLM-based                           & \num{34.38596491}                                      & {\num{46.4}}                          & \num{28.83333333}                          & \num{27.27272727}                          \\

        \bottomrule
    \end{tabular}
    \caption{
        Pass@$1$ performance on the LiveCodeBench test output prediction benchmark (Jan 1–Apr 1, 2024), {where difficulty levels follow the original LiveCodeBench setting based on the problem's difficulty}. Functional majority voting is applied to all baselines by default. 
    }
    \label{tab:cutoff_orig_split}
\end{table*}

%% file: table/code_filtering_humanevalplus.tex
\begin{table*}[htb!]
    \centering
    \small
    \renewcommand{\arraystretch}{1.2}
    \arrayrulecolor{black}
    \begin{tabular}{>{\color{black}}l >{\color{black}}l >{\color{black}}l >{\color{black}}c >{\color{black}}c >{\color{black}}c >{\color{black}}c >{\color{black}}c}
        \toprule
        \multirow{2}{*}{\textbf{Method}}  & \multirow{2}{*}{\textbf{Grounding}} & \multirow{2}{*}{\textbf{Execution}} & \multicolumn{2}{c}{\multirow{1}{*}{{\textbf{Pass@$1$}}}} \\
        & & & \multirow{-1.1}{*}{\textbf{HumanEval}} & \multirow{-1.1}{*}{\textbf{HumanEval+}} \\
        \hline
        \small No Filtering & -                                   & -                                   & \num{79.2}   & {69.9}          \\
        \small \textsc{CodeT}  & -                                   & -                                   & {84.7}  & {75.0}          \\
        \small TestChain   & Code                                & Direct                                  & {83.0}  & {73.6}           \\
        \small \ours       & Code \& Pseudocode                  & Direct \& LLM-based                 &  \textbf{85.1}   &  \textbf{75.4}           \\
        \small \phantom{0}- \textit{direct code exec.}   & Pseudocode                          & LLM-based                           &  84.8   & 75.2   \\
        \bottomrule
    \end{tabular}
    \caption{
    {End-to-end code generation performance of \texttt{Llama-3.1-70B-Instruct} under different test output prediction methods, evaluated on HumanEval and HumanEval+.}
    }
    \arrayrulecolor{black}
    \label{tab:code_filtering_humanevalplus}
\end{table*}

%% file: table/cutoff_exaone.tex
\begin{table}[t]
    \begin{flushleft}
    \small
    \renewcommand{\arraystretch}{1.1}
    \resizebox{1.0\linewidth}{!}{
    \begin{tabular}{>{\color{black}}l >{\color{black}}l >{\color{black}}l >{\color{black}}c >{\color{black}}c >{\color{black}}c}
        \arrayrulecolor{black}\toprule
        \textbf{Method}          & \textbf{Ground.} & \textbf{Exec.} & {\textbf{Pass@$1$} }                                                                                                              \\
        \hline
        \small \texttt{GPT-4-Turbo-2024-04-09}   & -                                   & -                                   & {\num{72.3684210526315}}\\
        \small TestChain            & C                                & D                              & {\num{74.34210526}} \\
        
        \small \textbf{\ours}                 & C \& P                  & D \& L                 & \textbf{\num{81.57894737}} \\
        \small \phantom{0}- \textit{direct code exec.}             & P                          & L                           & {\num{69.07894737}} \\
        \hline
        \small \texttt{Llama-3.1-8B-Inst}    & -                                   & -                                   & {\num{34.01315789}} \\
        \small TestChain             & C                                & D                              & {\num{46.27192982}} \\
        \small \textbf{\ours}                 & C \& P                  & D \& L                 & \textbf{{\num{53.28947368}}} \\
        \small \phantom{0}- \textit{direct code exec.}             & P                          & L                           & {\num{34.38596491}} \\
        \hline
        \small \texttt{Llama-3.1-70B-Inst}   & -                                   & -                                   & {\num{64.47368421}} \\
        \small TestChain             & C                                & D                              & {\num{71.49122807}} \\
        \small \textbf{\ours}                 & C \& P                  & D \& L                 & \textbf{{\num{78.28947368}}} \\
        \small \phantom{0}- \textit{direct code exec.}             & P                          & L                           & {\num{66.66666667}} \\

        \hline
        \small \texttt{Mistral-Large}            & -                                   & -                                   &  {\num{59.21052632}} \\
        \small TestChain             & C                                & D                              & {\num{55.70175439}} \\
        \small \textbf{\ours}                 & C \& P                  & D \& L                 & \textbf{{\num{67.10526316}}} \\
        \small \phantom{0}- \textit{direct code exec.}             & P                          & L                           & {\num{46.31578947}} \\

        \hline
        \small \texttt{\color{red!50} EXAONE-3.5-7.8B-Inst} & -                                   & -                                   & {\num{47.14912281}} \\
        \small TestChain             & C                                & D                              & {\num{45.70175439}} \\
        \small \textbf{\ours}                 & C \& P                  & D \& L                 & \textbf{{\num{59.64912281}}} \\
        \small \phantom{0}- \textit{direct code exec.}             & P                          & L                           & {\num{45.61403509}} \\

        \hline
        \small \texttt{\color{red!50}EXAONE-3.5-32B-Inst}  & -                                   & -                                   & {\num{50.98684211}} \\
        \small TestChain             & C                                & D                              & {{\num{59.86842105}}} \\
        \small \textbf{\ours}                 & C \& P                  & D \& L                 & \textbf{{\num{65.13157895}}} \\
        \small \phantom{0}- \textit{direct code exec.}             & P                          & L                           & {\num{51.09649123}} \\

        \arrayrulecolor{black}\bottomrule
    \end{tabular}}
    \end{flushleft}
    \caption{
        \edit{Pass@$1$ comparison on the LiveCodeBench test output prediction benchmark (Jan 1–Apr 1, 2024). Functional majority voting is applied to all baselines by default. {Each method is labeled by its grounding type, code (C) or pseudocode (P), and execution type, direct (D) or LLM-based (L).} Red-colored models have the possibility of data contamination.}
    }
    \label{tab:cutoff_exaone}
\end{table}

%% file: table/consensus_exaone.tex
\begin{table}[tb!]
    \centering
    \small
    \renewcommand{\arraystretch}{1.2}
    \resizebox{1.0\linewidth}{!}{
    \begin{tabular}{>{\color{black}}l >{\color{black}}l >{\color{black}}c >{\color{black}}c >{\color{black}}c}
        \arrayrulecolor{black}\toprule
        \textbf{Model} & \textbf{\ours Voting Method}                                                                            & \textbf{Pass@$1$}            \\
        \hline
        \small \multirow{2}{*}{\texttt{GPT-4-Turbo-2024-04-09}} & \textit{functional majority voting} & \num{81.6}       \\
        &  + \textit{path-weighted}      & \num{81.6}    \\
        \hline
        \small \multirow{2}{*}{\texttt{Llama-3.1-8B-Inst}} & \textit{functional majority voting}  & \num{53.3}       \\
        & + \textit{path-weighted}      & \num{53.3}    \\
        \hline
        \small \multirow{2}{*}{\texttt{Llama-3.1-70B-Inst}} & \textit{functional majority voting}  &  \num{77.6315789473684}      \\
        &  + \textit{path-weighted}     & \textbf{\num{78.3}}    \\
        \hline
        \small \multirow{2}{*}{\texttt{Mistral-Large}} & \textit{functional majority voting}      & \num{67.1052631578947}    \\
        & + \textit{path-weighted}  & \num{67.1}       \\
        \hline
        \small \multirow{2}{*}{\color{red!50}\texttt{EXAONE-3.5-7.8B-Inst}} & \textit{functional majority voting}      & \num{59.6491228070175}    \\
        & + \textit{path-weighted}  & \num{59.6}       \\
        \hline
        \small \multirow{2}{*}{\color{red!50}\texttt{EXAONE-3.5-32B-Inst}} & \textit{functional majority voting}      & \num{64.4736842105263} \\
        & + \textit{path-weighted}  & \textbf{\num{65.1}}       \\
        \arrayrulecolor{black}\bottomrule\arrayrulecolor{black}
    \end{tabular}}
    \caption{
        \edit{The effect of path-weighted FMV on the LiveCodeBench test output prediction benchmark (Jan 1–Apr 1, 2024). Red-colored models have the possibility of data contamination.}
    }
    \arrayrulecolor{black}
    \label{tab:consensus_exaone}
\end{table}

%% file: table/model_cutoff_info.tex
\begin{table}[t]
    \centering
    \small
    \arrayrulecolor{black}
    \begin{tabular}{>{\color{black}}l|c}
        \thickhline
        \multirow{2}{*}{\textbf{Model Name}} & \textbf{Approximate} \\
                                             & \textbf{Cutoff Date} \\
        \hline
        \texttt{GPT-4-Turbo-2024-04-09  }    & 2023-04-30           \\
        \texttt{Llama-3.1-8B-Instruct   }    & 2023-12-31           \\
        \texttt{Llama-3.1-70B-Instruct  }    & 2023-12-31           \\
        \texttt{Mistral-Large           }    & 2023-01-01           \\
        \texttt{EXAONE-3.5-7.8B-Instruct}    & 2024-11-30           \\
        \texttt{EXAONE-3.5-32B-Instruct }    & 2024-11-30           \\
        \thickhline
    \end{tabular}
    \caption{Cutoff dates for each LLM.}
    \label{tab:model_cutoff_dates}
\end{table}